\documentclass[preprint,12pt]{elsarticle}

\usepackage{amssymb}
\usepackage{amsmath}
\usepackage{graphicx}
\usepackage{booktabs}
\usepackage{longtable}
\usepackage{enumitem}
\usepackage{tabularx}
\usepackage{pdflscape}
\usepackage{siunitx}
\usepackage{xcolor}
\usepackage{hyperref}
\usepackage{amsmath,amssymb}
\usepackage{multirow}
\usepackage{cleveref}
\newtheorem{remark}{Remark}

\begin{document}

\begin{frontmatter}

\title{Are Large Language Models able to Predict  Highly Cited Papers? Evidence from Statistical Publications}

\author[1]{Zhanshuo Ye\fnref{fn1}}
\author[1]{Yiming Hou\fnref{fn1}}
\author[1]{Rui Pan\fnref{fn1}}
\author[2]{Tianchen Gao\corref{cor1}}
\author[3]{Hansheng Wang}

\affiliation[1]{
    organization={School of Statistics and Mathematics, Central University of Finance and Economics},
    country={China}
}

\affiliation[2]{
    organization={Beijing International Center for Mathematical Research (BICMR), Peking University},
    country={China}
}
\affiliation[3]{
    organization={Guanghua School of Management, Peking University},
    country={China}
}



\cortext[cor1]{Corresponding author: gaotc@pku.edu.cn}
\fntext[fn1]{These authors are co-first authors and contributed equally.}

\begin{abstract}
Predicting highly-cited papers is a long-standing challenge due to the complex interactions of research content, scholarly communities, and temporal dynamics. Recent advances in large language models (LLMs) raise the question of whether early-stage textual information can provide useful signals of long-term scientific impact. Focusing on statistical publications, we propose a flexible, text-centered framework that leverages LLMs and structured prompt design to predict highly cited papers. Specifically, we utilize information available at the time of publication, including titles, abstracts, keywords, and limited bibliographic metadata.
Using a large corpus of statistical papers, we evaluate predictive performance across multiple publication periods and alternative definitions of highly cited papers. The proposed approach achieves stable and competitive performance relative to existing methods and demonstrates strong generalization over time. Textual analysis further reveals that papers predicted as highly cited concentrate on recurring topics such as causal inference and deep learning. To facilitate practical use of the proposed approach, we further develop a WeChat mini program, \textit{Stat Highly Cited Papers}, which provides an accessible interface for early-stage citation impact assessment. Overall, our results provide empirical evidence that LLMs can capture meaningful early signals of long-term citation impact, while also highlighting their limitations as tools for research impact assessment. 

\end{abstract}





\end{frontmatter}

\newpage

\section{Introduction}

The ability to evaluate and predict the long-term scholarly impact of scientific publications is a central theme in research evaluation. Citation counts, in particular, is the most widely used and visible indicator of academic influence \cite{waltman2016}. They often serve as a benchmark for research quality, visibility and prestige, and are valued because they are easy to quantify and can be compared between disciplines when properly normalized \cite{waltman2013percentiles}.
As a result, funders, tenure committees, and institutions increasingly rely on citation-based indicators in their decisions \cite{hicks2015}.
However, citation accumulation is typically slow.
Many influential papers take years to gain broad recognition, creating a clear time lag between publication and impact \cite{wang2013}. Some groundbreaking works may not reach their citation peak until a decade or more after publication. This creates a clear time lag that makes real-time assessment particularly difficult.
Moreover, citation distributions are highly skewed \cite{seglen1992}.
Most papers receive few citations, while a small fraction achieve exceptional visibility and become highly cited \cite{seglen1992,ioannidis2022}.
This long-tail distribution illustrates both the value and limitations of the citation-based assessment \cite{waltman2016}. As a result, early and timely identification of such highly cited articles remains highly challenging but critically important \cite{hu2023_ipm}.

A wide range of studies have attempted to address this challenge by identifying factors that drive citation impact and by developing models to predict future highly cited papers. Traditional approaches generally rely on bibliometric and contextual features such as author reputation \cite{kosteas2018predicting}, journal impact \cite{traag2021inferring}, collaboration patterns \cite{gazni2014long}, and textual attributes of titles and abstracts \cite{dorta2018characterizing}. These predictors, combined with regression or machine learning models, have yielded valuable insights into the mechanisms of scientific influence \cite{hu2020identification}. However, these approaches typically require extensive feature engineering, rely heavily on manually designed features and structured paper data, and often fail to capture the semantic information embedded in the text. In terms of modeling, they also demand the design and tuning of various complex models, such as neural networks \cite{zhang2024_scientometrics}. Moreover, their predictive performance can vary across different disciplinary contexts \cite{zhang2024_scientometrics}. 
As a result, the prediction of highly cited papers remains both practically important and methodologically difficult.

The recent emergence of large language models (LLMs) such as ChatGPT, Gemini, and DeepSeek provides new opportunities to advance the prediction of scientific impact \cite{huang2025identifying}. These models demonstrate remarkable abilities in language understanding, reasoning, and capturing contextual information in diverse domains \cite{openai2023gpt4,deepseekV3_2024,gemini2024,deepseekR1_nature_2025}. In the field of text analysis, LLMs achieve notable success, including information extraction, summarization, topic modeling, sentiment analysis, and named entity recognition. Their applications span various fields, including medicine \cite{xie2025medical}, education \cite{parker2025large}, and social sciences \cite{deepseekR1_nature_2025}. Recently, LLMs have shown great potential in predicting scientific impact by capturing early textual signals of novelty \cite{lin2025schnovel,huang2025large}, forecasting citation potential \cite{zhao2025newborn}, and integrating semantic and network information \cite{hao2024hlmcite}. We argue that LLMs can process the text of scientific papers. Using titles, abstracts, and keywords, they learn semantic patterns that indicate novelty, contribution, and potential influence. Therefore, this study is motivated to use LLMs to predict highly cited papers, with the aim of exploring their ability to identify impactful research at an early stage.

To this end, our analysis is based on a large-scale dataset of 90,167 statistical papers published between 1991 and 2023 \cite{gao2023largescalemultilayeracademicnetworks}. Specifically, each paper includes bibliographic and textual information such as title, publisher, publication year, abstract, and keywords. To account for citation lags, we adopt the widely recognized percentile threshold method from bibliometrics, defining highly cited papers as those ranked within the top $1\%$, $5\%$, and $10\%$ by citation counts in the same year \cite{waltman2016}. Based on this dataset, we develop a structured evaluation framework using prompts to adapt the task of predicting high-cited papers by LLMs. This framework simulates the domain expert judgment process through a comprehensive prompt design with five core components. It requires LLMs to evaluate each paper based on three key criteria, i.e., methodological innovation, long term impact, and importance of the problem. The framework uses historical research trends and positive/negative examples to keep the evaluation consistent over time and reliable. We conduct a fair evaluation of three mainstream LLMs, i.e., ChatGPT, Gemini and DeepSeek. We analyze their accuracy (ACC), true positive rate (TPR), false positive rate (FPR), computational efficiency, and consistency over different time spans. The goal is to assess the potential of LLMs in identifying highly impactful research in the early stages and to provide insights into their utility as tools to predict scientific trends.

The main contributions of this study are threefold. First, using an existing large-scale dataset of statistical publications, we conduct a systematic investigation into whether the future influence of a paper can be inferred from early-stage textual information. This analysis formulates a well-defined empirical setting in which LLMs are applied to the prediction of highly cited papers in statistical science, providing evidence on the feasibility and limitations of text-based early impact assessment. 
Second, we propose a structured prompt engineering framework that transforms citation prediction into a format suitable for LLM-based reasoning. This framework enables a systematic comparison of different LLMs in terms of predictive accuracy, computational efficiency, and consistency across tasks. 
Third, by analyzing the papers that LLMs repeatedly identify as high-impact candidates, we uncover emerging research topics and evolving methodological directions in statistics. These findings demonstrate that LLM-based prediction not only offers a complementary tool for assessing scientific impact, but also provides insights that support research trend analysis and knowledge discovery.

The remainder of this work is organized as follows. Section 2 reviews the related literature and provides an overview of existing approaches. In Section 3, we present the data collection process and descriptive analysis. Our LLM-powered prediction framework is introduced in Section 4, including prompt engineering and LLM interaction. Section 5 reports the highly cited paper prediction results, including overall performance, comparisons across different LLMs and the stability of LLMs. The identification of future research directions is discussed in Section 6, with concluding remarks provided in Section 7.


\section{Literature Review}

This section reviews the existing literature related to the prediction of scientific impact and highly cited papers. We first summarize previous studies on predictors of citation impact, covering bibliometric, contextual, and network-based factors that influence the long-term visibility of articles. Second, we review the main modeling approaches developed for citation prediction, ranging from traditional regression and classification models to more recent ensemble and deep learning frameworks. Finally, we discuss emerging LLM-based methods that leverage semantic understanding of scientific texts to predict research impact, highlighting their advantages, limitations, and potential integration with conventional bibliometric indicators.

\subsection{Predictors of Citation Impact}
Existing work on predicting highly cited articles uses mainly bibliometric and context features \cite{gao2024}. These bibliometric features are generally categorized into three levels, including the author level, the journal level, and the network level. At the author level, predictors capture individual reputation and productivity. Traditional metrics include the $h$-index \cite{hirsch2005}, $g$-index \cite{du2019understanding} and field-normalized indicators \cite{waltman2015field}.
Recent work incorporates temporal patterns of author productivity, showing that career trajectories can improve the prediction of citations \cite{acuna2012predicting}.
Beyond productivity metrics, indicators of academic prestige also play an important role. These include team size, institutional affiliation, funding record, and international collaboration \cite{hirsch2005,tahamtan2019}.
At the journal level, predictors reflect the characteristics of the publication venue. Common indicators include the journal impact factor and the subject categories \cite{waltman2016,jci2021}. Recent studies examine alternative metrics such as CiteScore, SCImago journal rank, and source normalized impact per article. These alternatives provide more robust cross-disciplinary comparisons \cite{waltman2013some}.
At the network level, predictors have attracted increasing attention in recent years. They capture structural patterns within the scientific community that traditional metrics ignore \cite{gao2024}. Co-authorship networks are particularly informative, as they reveal collaborative advantages through centrality measures. These measures include degree centrality, betweenness centrality, and closeness centrality \cite{sarigol2014predicting}. Similarly, co-citation networks capture intellectual similarity and knowledge flow patterns. Studies demonstrate that papers frequently co-cited together exhibit thematic relatedness that correlates with future impact \cite{trujillo2018document}. Moreover, citation network dynamics proves to be effective for early prediction of paper impact. Metrics such as early citation velocity and the diversity of citation papers can identify potential breakthrough papers shortly after publication \cite{ponomarev2014predicting}.

In addition to these bibliometric features, context features provide complementary information for citation prediction, especially when early citation data are unavailable \cite{hu2020identification}. These features include content-based predictors such as titles, abstracts, keywords, and publication metadata.
Specifically, title features significantly influence citations. Shorter, declarative titles that present the main results tend to receive higher citations. The characteristics of the titles vary between disciplines. Disciplines such as physics and chemistry tend to emphasize methodological approaches in titles, while disciplines such as sociology and psychology more often highlight research findings \cite{milojevic2017length}.
Beyond titles, abstract features are particularly effective in capturing semantic content and linguistic characteristics that effectively reflect research quality and predict citations \cite{dorta2018characterizing}. Embedding techniques from TF-IDF to GPT models extract topics, writing style, semantics, and sentiment from abstracts \cite{zhang2024_scientometrics}. Additionally, semantic indices quantify stylistic features such as semantic span and inflection, providing complementary information for citation prediction \cite{zhu2024instant}.
Keywords also play a role in citation prediction by indicating main research topics and field visibility \cite{dorta2018characterizing}, thus improving prediction accuracy \cite{la2021leveraging}. 
Another important content-based predictor is journal information, which serves as an important contextual factor, as venues differ in visibility and citation behavior across disciplines \cite{traag2021inferring}.
In summary, combining bibliometric and contextual features, especially with early citation and altmetric data, substantially improves long-term impact prediction \cite{hu2020identification,gao2024}.

\subsection{Citation Impact Prediction Models}

To make full use of both bibliometric and contextual features, researchers develop various models to predict citation counts and identify influential papers. These models are generally divided into regression and classification approaches, depending on the form of the target variable. 
Regression approaches treat citation counts as continuous variables and aim to predict exact citation numbers. Early linear regression models \cite{bornmann2014improve} struggle with the non-negative, skewed, and zero-inflated nature of citation data, leading to widespread adoption of Poisson and negative binomial regressions \cite{ajiferuke2015modelling,onodera2015factors}. Extensions include quantile regression for distribution-varying effects \cite{stegehuis2015} and Bayesian hurdle models for zero-inflation \cite{shahmandi2021bayesian}.
In contrast, classification approaches solve the problem by categorizing papers according to their relative citation performance using pre-defined thresholds, such as top percentiles within a field or time period. 
Among classification methods, logistic regression  provides interpretable baselines for binary high-impact classification \cite{marques2024ten}, while support vector machines handle high-dimensional features and capture nonlinear patterns through kernel functions \cite{velkumar2025improving}.

Although traditional regression and classification models provide interpretable baselines, they face limitations when dealing with complex citation patterns. These methods rely on manual feature engineering and linear or traditional statistical models, which struggle to capture the nonlinear relationships that shape citation behavior \cite{zhang2024_scientometrics}. To address these constraints, researchers are increasingly adopting deep learning approaches for predicting citation counts. Among these, deep neural networks offer several advantages. They automatically learn hierarchical representations directly from raw data, eliminating the need for extensive manual feature engineering. In addition, they model nonlinear dependencies among various feature types, including bibliometric indicators, textual content, and network structures. 

Various deep learning architectures leverage these capabilities in different ways \cite{zhang2024_scientometrics,liang2021combining,huang2022fine}. Convolutional neural networks(CNNs) capture nonlinear interactions between early network features and later citation outcomes, effectively modeling complex dependencies among publication attributes \cite{zhang2024_scientometrics}. To better represent the evolving nature of citation behavior, hybrid frameworks such as DeepCCP combine GRU and CNN modules, enabling dynamic modeling of early citation cascades \cite{zhao2022utilizing}. Graph neural networks use both node features and citation links to represent the structure of citation networks \cite{cummings2020structured}. Meanwhile, recurrent neural networks and long short-term memory architectures focus on capturing temporal patterns in citation growth, forming the task as a time-series prediction problem \cite{abrishami2019predicting}.

\subsection{LLM-based Approaches}

With the advancement of LLMs, the research of citation prediction shifts from feature-engineered machine learning toward text driven and knowledge enhanced approaches. LLMs offer several key advantages over traditional methods.
First, they extract meaningful signals from early-stage text. This capability is in accordance with expert assessments of research novelty through various prompting strategies \cite{huang2025papereval,lin2025schnovel}. For instance, systems such as GraphMind support novelty evaluation by building knowledge graphs and comparing papers against related work \cite{da2025graphmind}. 
Second, LLMs demonstrate a strong ability to understand citation contexts. They capture the semantic meanings of the citation contexts and generate appropriate citations for given passages \cite{zhang2023large}. Frameworks such as ALCE evaluate citation quality in multiple aspects, including relevance, ACC, and completeness \cite{gao2023enabling}. 
Third, LLMs treat impact prediction as a text comprehension task. Rather than relying on engineered features, they predict citations and engagement directly from paper abstracts. Recent studies show that title-abstract models produce competitive forecasts even for newly published papers without prior citation history \cite{zhao2025newborn}.

Despite these promising capabilities, LLM-based approaches still face several challenges.
First, they exhibit bias in citation behavior. When asked to recommend citations for a given topic or context, models like GPT 4o or Claude 3.5 tend to suggest highly cited papers. This occurs even when less-cited papers may be equally relevant or more appropriate for the specific context. Such bias can strengthen the Matthew effect, where popular papers continue to gain more citations while less-cited papers remain overlooked. 
Second, predictive performance differs between disciplines and tasks. Cross-disciplinary evaluations reveal that the ability of ChatGPT to generate citations and references varies substantially. This variation is influenced by factors such as DOI availability standards and field-specific citation conventions \cite{algaba2024large}. In addition, some prediction tasks are more challenging than others. Predicting citation counts is generally more difficult than assessing readability or novelty \cite{zhao2025newborn}.
Third, technical constraints limit practical implementation. Context window size restricts the amount of information that the models can process at once. Additionally, LLMs rely heavily on their training data, which may not reflect the latest research developments. The black-box nature of these models also poses challenges. It reduces reproducibility and makes it difficult to understand how predictions are generated \cite{algaba2024large}.

To address these limitations, recent work explores hybrid approaches that combine the strengths of LLMs with traditional bibliometric methods. These frameworks utilize the complementary strengths of both approaches. First, integrating LLM reasoning with citation-based retrieval is effective. This integration identifies highly cited citations more accurately than using either method alone \cite{hao2024hlmcite}. Similarly, combining semantic features extracted by LLMs with traditional bibliometric features improves overall prediction accuracy \cite{zhang2024_scientometrics}.
Second, retrieval-augmented generation offers important benefits for citation prediction tasks. This approach improves LLM performance by incorporating external data sources during generation \cite{li2026towards}. Moreover, graph retrievers enable LLMs to better understand the co-citation structure by providing structured knowledge \cite{hao2024hlmcite}.
Third, specialized frameworks demonstrate the value of adaptive learning techniques. Studies show that in-context learning and fine-tuning enhance the performance of
LLMs, especially when models incorporate citation context features. These models substantially outperform the baseline approaches \cite{hao2024hlmcite}.
Despite these advances, systematic frameworks that combine LLMs with bibliometric features for impact prediction remain underdeveloped. This gap presents an important direction for future research.

\section{Data Collection and Descriptive Analysis}

\subsection{Data Collection}

This study uses a dataset comprising publications from 42 representative journals in the field of statistics \cite{gao2023largescalemultilayeracademicnetworks}. These journals are selected based on their classification in the Journal Citation Reports (JCR), with a primary focus on the categories of {\it Statistics \& Probability} and {\it Economics}, including additional journals in Biostatistics, Machine Learning, and Data Mining. The data are gathered from the {\it Web of Science} (\url{www.webofscience.com}), an authoritative academic database, and include 90,167 papers published between 1991 and 2023. Each paper recorded in the dataset contains title, publisher, abstract, keywords, publication year and citation counts up to August 2025, as shown in Table~\ref{tab:paper-example}. Specifically, the variable of citation counts represents the cumulative number of citations recorded in {\it Web of Science} up to August 2025. 

\begin{table}[!ht]
\centering
\caption{An illustrative example of a paper in the dataset}
\label{tab:paper-example}
\small 
\begin{tabular}{@{} l p{0.75\textwidth} @{}} 
\toprule
\textbf{Variable} & \textbf{Example} \\
\midrule
Title & Controlling the False Discovery Rate: A Practical and Powerful Approach to Multiple Testing \\
Authors & Yoav Benjamini; Yosef Hochberg \\
Journal & Journal of the Royal Statistical Society: Series B - Statistical Methodology \\
Abstract & The common approach to the multiplicity problem calls for controlling the familywise error rate (FWER). This approach, though, has faults, and we point out a few. A different approach to problems of multiple significance testing is presented. It calls for controlling the expected proportion of falsely rejected hypotheses - the false discovery rate. This error rate is ... \\
Keywords & bonferroni-type procedures, familywise error rate, multiple-comparison procedures, p-values \\
Year & 1995 \\
Citation Counts & 92,564 (up to August 2025) \\
\bottomrule
\end{tabular}
\end{table}

\subsection{Descriptive Analysis}

\begin{figure}[t]
    \centering
    \includegraphics[width=1\linewidth]{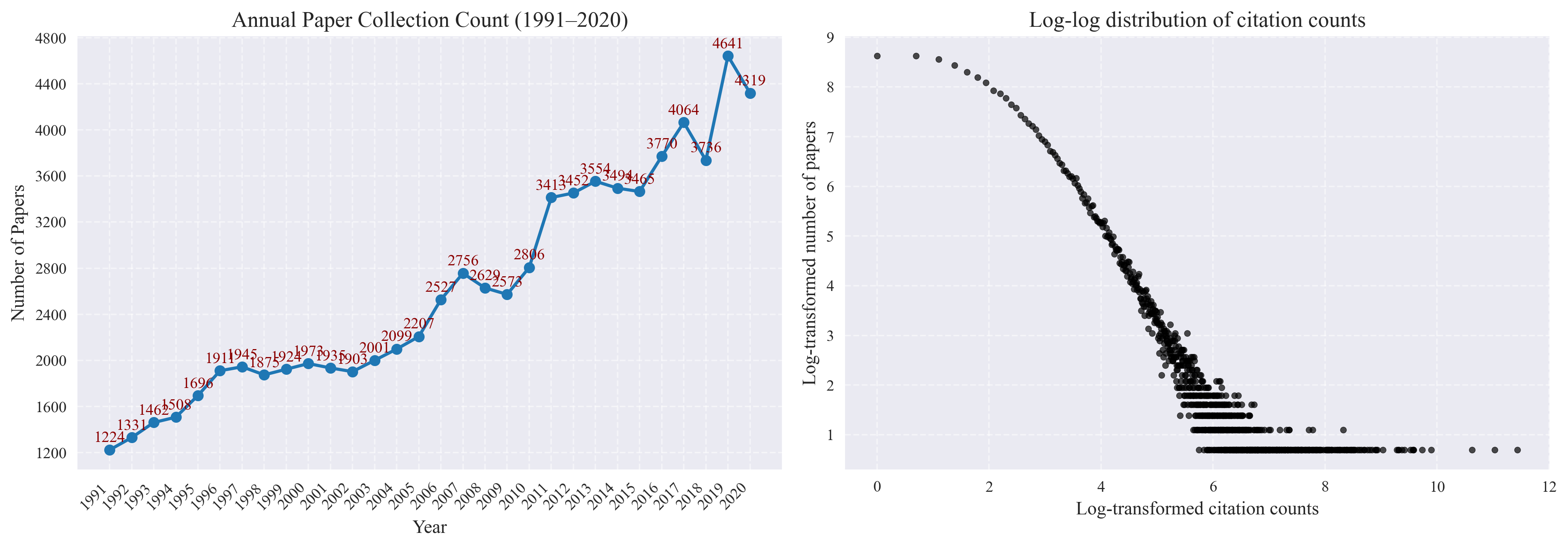}
    \caption{Left: Annual number of papers collected by our research group. The series exhibits a clear upward trend, with particularly rapid growth after 2009. Right: Log–log distribution of citation counts, showing a pronounced long-tailed pattern that approximately follows a power-law distribution.}   \label{fig:papers_per_year_citation_distribution}
\end{figure}

The left plot in Figure~\ref{fig:papers_per_year_citation_distribution} presents the annual number of papers collected in our dataset. From 1991 to the early 2000s, the annual number of papers rises moderately, maintaining a relatively stable range between 1,200 and 2,000 publications per year. Starting around 2005, the number of publications begins to accelerate, with a sharp increase observed after 2009. This surge corresponds to the broader growth of data-driven research and the interdisciplinary integration of statistics with fields such as economics, biomedicine, and data mining. The right plot displays the log–log distribution of citation counts, which reveals a long-tailed pattern and this pattern roughly follows a power-law distribution {\cite{seglen1992,redner1998}}. Specifically, 6,229 papers (6.9\%) have never been cited, and 31,231 papers (34.6\%) have received no more than five citations. In contrast, a few highly influential papers have accumulated tens of thousands of citations. For example, the most cited paper is ``Controlling the False Discovery Rate: A Practical and Powerful Approach to Multiple Testing", published in {\it Journal of the Royal Statistical Society: Series B} (1995). It has received 92,564 citations, reflecting its enduring impact in the field of statistics. The above observations provide further evidence for the earlier claim that citation impact is highly skewed. This makes it challenging to model or predict citation counts, because most papers have low impact while only a few receive very high numbers of citations.

\begin{table*}[ht]
\centering
\caption{Summary statistics for the top 5\% of citation counts across publication groups from 1991 to 2020. Reported metrics include the mean, median, standard deviation (SD), interquartile range (IQR), and the maximum observed citation counts.}
\label{tab:top5_summary}
\small
\begin{tabular}{cccccc}
\toprule
\textbf{Publication Group} & \textbf{Mean} & \textbf{Median} & \textbf{SD} & \textbf{IQR} & \textbf{Max} \\
\midrule
        1991-1995 & 944.6  & 358  & 4792.9  & 367  & 92564   \\ 
        1996-2000 & 637.5  & 304  & 1360.4  & 308  & 19826   \\ 
        2001-2005 & 510.0  & 278  & 933.6  & 263  & 14482   \\ 
        2006-2010 & 531.3  & 240  & 1102.8  & 270  & 13690   \\ 
        2011-2015 & 301.5  & 134  & 2130.1  & 119  & 61611   \\ 
        2016-2020 & 153.1  & 76  & 651.1  & 59 & 16880   \\ 
\bottomrule
\end{tabular}
\end{table*}

Next, we provide a descriptive overview of the top 5\% of citations within each publication year. In the subsequent analysis, we extend the evaluation to alternative thresholds (e.g., 10\%) to ensure a more comprehensive comparison. Table~\ref{tab:top5_summary} shows the yearly summary statistics for the top 5\% most cited papers from 1991 to 2020. Even within this highly cited group, citation impact remains strongly heterogeneous. Two clear patterns appear in the summary statistics. First, citation counts tend to decrease for more recent publication groups, as reflected in the decreasing means and medians. Second, both SD and IQR become smaller with time, meaning that the spread of citation impact has narrowed. These summary statistics clarify how citation impact varies across groups and illustrate the evolving influence of highly cited papers over time. This internal heterogeneity, combined with the rarity of extremely influential papers, highlights the challenge of identifying truly high-impact work.

\section{LLM-powered Prediction Framework}

\begin{figure}[!ht]
    \centering
    \includegraphics[width=0.75\linewidth]{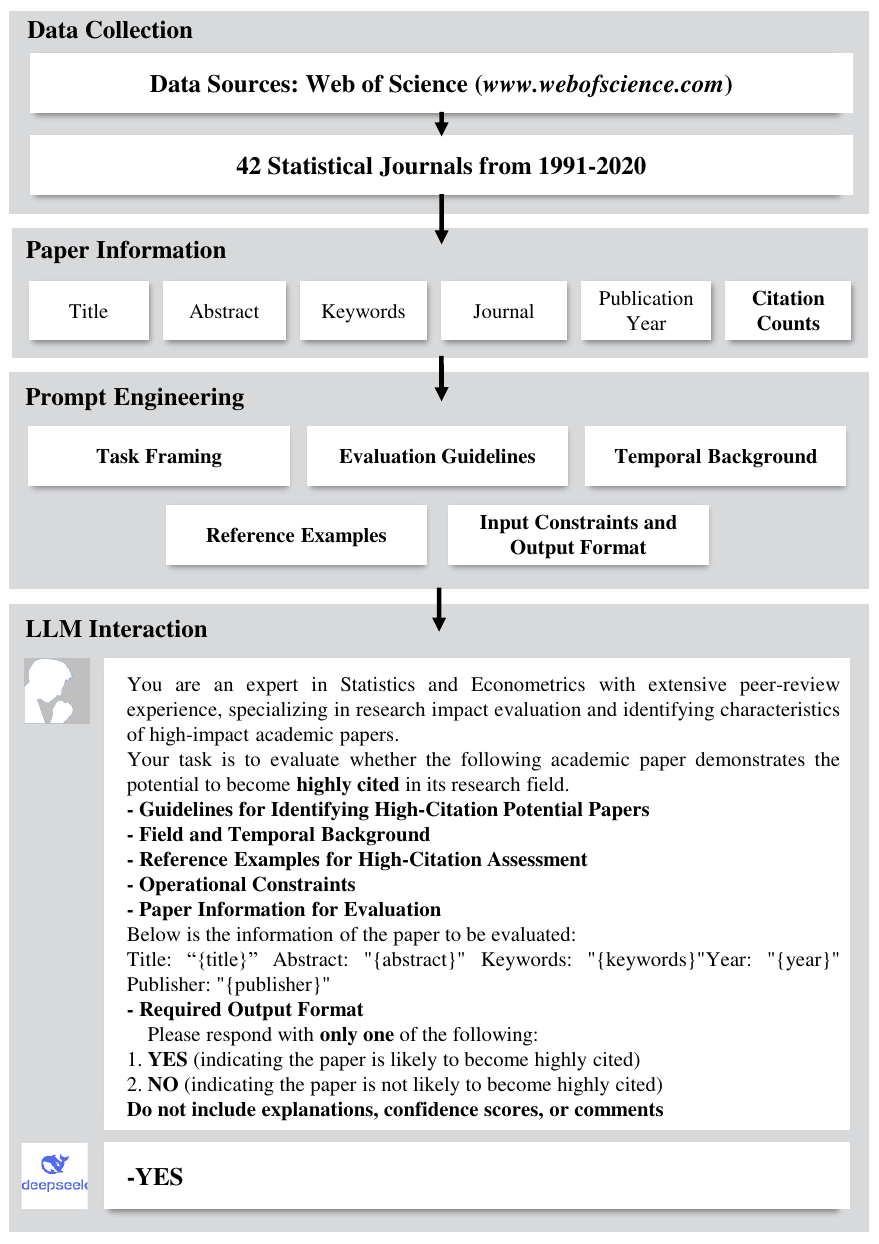}
    \caption{Overall methodological framework integrating data collection, prompt design, and predictive modeling by LLM interaction.}
    \label{fig:framework}
\end{figure}

This section outlines the general methodological framework. Figure~\ref{fig:framework} summarizes the workflow and shows how bibliometric data are integrated with text-based predictive modeling.
It further highlights the use of structured prompts and LLM interaction within the proposed approach.

\subsection{Prompt Engineering}

To adapt the prediction task for LLMs, we develop structured prompts that explicitly define the role of the model, evaluation criteria, temporal background, reference examples, and input–output format. The analysis uses the collected citation counts as the true labels for evaluation, and the prompt design specifies how the model interprets the textual and bibliographic information available at publication. Together, these elements create a controlled setting in which the predictive capabilities of LLMs can be systematically examined. Within this setting, the model is asked to determine whether a paper is in the top $k\%$ of citations with $k = 1, 5, 10$. The framework for this task consists of five components, which are described below.

\noindent
\textbf{Task Framing.} The prompt defines the role of the model as an expert in statistics and econometrics. This matches the scope of our dataset, which covers core journals in these fields, and ensures that the evaluations are situated within the proper academic context. 
The model is instructed to assess whether a given paper has the potential to become highly cited. We do not provide the specific definition of highly cited papers to prevent the LLMs from relying on external citation information or post-publication data. As a result, the judgment is based solely on the textual and contextual information provided in the prompt.

\noindent
\textbf{Evaluation Guidelines.} To define clear criteria for identifying potentially highly cited papers, we gather aims and scopes from four flagship statistical journals, i.e., \textit{Annals of Statistics}, \textit{Biometrika}, \textit{Journal of the American Statistical Association} and \textit{Journal of the Royal Statistical Society: Series B}. We also include aims and scopes from six randomly selected journals. Using LLMs, we summarize the textual content of the journals into three main categories, including methodological innovation, long-term value, and the significance of the problem.

\noindent 
\textbf{Temporal Background.} Since methodological innovation evolves over time \cite{dhingra2022}, the prompts include background summaries for each five-year period from 1991 to 2020, describing major research trends and developments in both methodological and applied areas. Specifically, when evaluating papers published within a given five-year window, the temporal background is generated using ChatGPT-4o mini by summarizing academic trends from the preceding five years. This background is constructed only from information available prior to the evaluation period and is kept fixed across all papers within the same time window. The temporal background is then included in the prompt to approximate the research context at the time of publication, allowing the model to assess potential impact based on contemporaneous information.

\noindent 
\textbf{Reference Examples.} Each prompt additionally provides six sample papers, including three highly cited papers from different journals as positive cases and three non-highly cited papers as negative cases. Each sample includes the title, abstract, keywords, and publisher, together with the true label. These examples help the model produce judgments consistent with the intended definition of high citation impact.

\noindent 
\textbf{Input Constraints and Output Format.} To ensure that evaluations rely only on information available at publication time and to prevent data leakage \cite{kaufman2012}, the model accesses only the title, abstract, keywords, publisher and year, without external databases or future citation records. The model outputs a binary classification, i.e., “YES” if a paper is likely to become highly cited and “NO” otherwise.

\begin{remark}[Potential Information Leakage]
A natural concern in applying LLMs to citation impact prediction is the possibility of information leakage, as contemporary LLMs may have been pretrained on corpora that extend beyond the publication period of the evaluated papers. We emphasize that our study does not aim to construct a strictly forward-looking or causally identifiable prediction system in the sense of training models on temporally truncated corpora. Instead, our objective is to examine whether, under explicitly controlled input constraints, LLMs can extract early-stage textual and contextual signals that are systematically associated with long-term citation impact.

In our design, each prompt corresponds to a single paper. The model is provided only with information available at the time of publication, including the title, abstract, keywords, journal, and year. The prompt does not disclose how highly cited papers are defined, nor does it specify any citation thresholds or reference to our database construction. As a result, even if an LLM were to recall or infer citation-related information, it would not be able to determine whether a paper meets the criteria for being highly cited in our study.

Under these constraints, model judgments must rely on patterns in research topics, problem formulation, methodological positioning, and scholarly language, rather than on explicit knowledge of future citation outcomes. We acknowledge that completely eliminating all forms of implicit temporal knowledge would require retraining or time-sliced pretraining of LLMs, which is a system-level modeling problem beyond the scope of this application-oriented study. Our results should therefore be interpreted as evidence of the extent to which LLMs can approximate expert-style assessments of potential citation impact based on contemporaneous textual information.
\end{remark}

\subsection{LLM Interaction}

Unlike traditional predictive methods that rely on statistical modeling, machine learning classifiers, or deep neural architectures, our approach uses LLMs directly as prediction engines. Instead of constructing complex feature representations or designing specialized model architectures, we rely on advanced LLMs. These LLMs interpret the textual content of a paper to assess whether it is likely to become highly cited. As a result, our approach eliminates the need for feature engineering or field-specific models. It relies only on the text available at the time of publication and frames the task as a semantic assessment rather than a statistical learning problem. This allows us to study the ability of LLMs to detect early signals of scholarly impact.
To apply this framework in practice, we interact with LLMs through their official Application Programming Interfaces (APIs). Each model (ChatGPT 4o mini, DeepSeek R1, DeepSeek V3, and Gemini 2.0 Flash) provides an API endpoint that accepts a prompt and returns a generated response. API usage follows a token-based billing structure, where costs depend on the number of input and output tokens processed during each request. For each paper in our dataset, a structured prompt is assembled programmatically using its title, abstract, keywords, publisher, and publication year. Tables~\ref{tab:prompt-full} and \ref{tab:prompt-examples} present a full example for the period 2001 to 2005; detailed templates for all five-year intervals are provided in Appendix~A.

\setlength{\LTpre}{8pt}
\setlength{\LTpost}{8pt}

\begin{scriptsize}
\renewcommand{\arraystretch}{1.1}
\begin{longtable}{@{} p{0.18\linewidth} p{0.76\linewidth} @{}} 
\caption{Structured prompt components (2001--2005)}
\label{tab:prompt-full}\\
\toprule
\textbf{Category} & \textbf{Prompt Content} \\
\midrule
\endfirsthead
\multicolumn{2}{c}{\tablename~\thetable{} -- continued} \\
\toprule
\textbf{Category} & \textbf{Prompt content} \\
\midrule
\endhead
\midrule
\multicolumn{2}{r}{\small Continued on next page} \\
\endfoot
\bottomrule
\endlastfoot

Task framing &
You are an expert in Statistics and Econometrics with extensive peer-review experience, specializing in research impact evaluation and identifying characteristics of high-impact academic papers.\par
Your task is to evaluate whether the following academic paper demonstrates the potential to become highly cited in its research field. \\ \midrule

Evaluation\par guidelines &
To guide your evaluation, consider the following characteristics of highly cited work:\par
\textit{1) Methodological Innovation}: proposes new methods or significantly improves existing ones; solves key bottlenecks (e.g., efficiency, ACC); connects methods with theory or practice.\par
\textit{2) Long-term Value}: advances foundational theories; has potential to shift paradigms.\par
\textit{3) Problem Significance}: addresses high-stakes scientific or societal problems; shows clear practical utility. \\ \midrule

Temporal\par background &
Papers published in the late 1990s (1996--2000) emerged during a period marked by accelerated methodological innovation and the consolidation of computationally intensive approaches. Advances in nonparametric regression, spline smoothing, and Bayesian hierarchical modeling gained adoption in Statistics. In Econometrics, increased availability of panel data stimulated work on dynamic panel estimators and generalized method of moments (GMM), while simulation-based inference expanded the practical analytical toolkit. Machine Learning was shaped by support vector machines, boosting, and ensemble methods, which offered new paradigms for prediction and classification. Computational statistics benefited from developments in MCMC, enabling Bayesian applications to high-dimensional problems. Applications in macroeconomics, finance, and social sciences were complemented by cross-disciplinary research in bioinformatics and information retrieval, reflecting the importance of large heterogeneous datasets.\\ \midrule

Reference\par examples &
Detailed illustrative examples (three highly cited and three not highly cited) are presented in Table~\ref{tab:prompt-examples}. Due to space limitations, abstracts of examples have been shortened for conciseness. \\ \midrule

Input constraints\par and output\par format &
\textit{Operational constraints}:\par
\begin{itemize}
  \item Base judgment only on: \textbf{Title, Abstract, Keywords, Publisher, Publication Year}.
  \item Consider the scientific landscape and prevailing trends at the time of publication (based on the year).
  \item Do not assume access to external databases or post-publication information.
\end{itemize}
\textit{Paper information for evaluation}:\par
Title: \{title\}\par
Abstract: \{abstract\}\par
Keywords: \{keywords\}\par
Year: \{year\_cleaning\}\par
Publisher: \{publisher\}\par
\medskip
\textit{Required output format}:\par
Respond with \textbf{only one} of the following:\par
-- \textbf{YES} (the paper is likely to become highly cited)\par
-- \textbf{NO} (the paper is not likely to become highly cited)\par
\medskip
Do not include explanations, confidence scores, or comments. \\
\end{longtable}
\end{scriptsize}

\begin{scriptsize}
\renewcommand{\arraystretch}{1.1}
\begin{longtable}{@{} p{0.98\linewidth} @{}} 
\caption{Illustrative reference examples (2001--2005)}
\label{tab:prompt-examples}\\
\toprule
\textbf{Examples} \\
\midrule
\endfirsthead
\multicolumn{1}{c}{\tablename~\thetable{} -- continued} \\
\toprule
\textbf{Examples} \\
\midrule
\endhead
\midrule
\multicolumn{1}{r}{\small Continued on next page} \\
\endfoot
\bottomrule
\endlastfoot

Below are several examples illustrating how to distinguish between highly cited and not highly cited papers.\par
\medskip

\textbf{Example 1: Highly Cited}\par
Title: \textit{Greedy function approximation: A gradient boosting machine}\par
Publisher: Annals of Statistics\par
Abstract: Function estimation/approximation is viewed from the perspective of numerical optimization in function space, rather than parameter space. A connection is made between stagewise additive expansions and steepest-descent minimization. A general gradient descent ``boosting'' paradigm is developed for additive expansions based on any fitting criterion... \par
Keywords: function estimation; boosting; decision trees; robust nonparametric regression\par
Judgment: \textbf{YES} \\
\addlinespace[1ex]

\textbf{Example 2: Highly Cited}\par
Title: \textit{Regularization and variable selection via the elastic net}\par
Publisher: Journal of the Royal Statistical Society: Series B (Statistical Methodology)\par
Abstract: We propose the elastic net, a new regularization and variable selection method. Real world data and a simulation study show that the elastic net often outperforms the lasso, while enjoying a similar sparsity of representation... \par
Keywords: grouping effect; LARS algorithm; Lasso; penalization; $p \gg n$ problem; variable selection\par
Judgment: \textbf{YES} \\
\addlinespace[1ex]

\textbf{Example 3: Highly Cited}\par
Title: \textit{Variable selection via nonconcave penalized likelihood and its oracle properties}\par
Publisher: Journal of the American Statistical Association\par
Abstract: Variable selection is fundamental to high-dimensional statistical modeling, including nonparametric regression. Many approaches in use are stepwise selection procedures, which can be computationally expensive and ignore stochastic errors in the variable selection process... \par
Keywords: hard thresholding; LASSO; nonnegative garrote; penalized likelihood; oracle estimator; SCAD; soft thresholding\par
Judgment: \textbf{YES} \\
\addlinespace[1ex]

\textbf{Example 4: Not Highly Cited}\par
Title: \textit{Profile quasi-likelihood}\par
Publisher: Statistics \& Probability Letters\par
Abstract: In this paper, the only assumptions on the distribution of data are those concerning first two moments. Our purpose is to estimate the parameter of interest in the presence of nuisance parameter under these weak assumptions on the distribution... \par
Keywords: quasi likelihood; profile quasi-likelihood; efficiency; semiparametric\par
Judgment: \textbf{NO} \\
\addlinespace[1ex]

\textbf{Example 5: Not Highly Cited}\par
Title: \textit{Bayes methods for outliers in binomial samples}\par
Publisher: Communications in Statistics -- Theory and Methods\par
Abstract: This article is concerned with the detection of outliers in a binomial sample. A Bayesian approach to the modeling of outliers is presented and examined. It is supposed that most observations are from a binomial distribution with mean $\pi$ but a small number of observations may be contaminated... \par
Keywords: Bayesian methods; binomial samples; outliers\par
Judgment: \textbf{NO} \\
\addlinespace[1ex]

\textbf{Example 6: Not Highly Cited}\par
Title: \textit{Comparing treatment strategies using a synthesized clinical trial: an analysis of late versus early use of trimethoprim-sulfamethoxazole for AIDS patients}\par
Publisher: Journal of Statistical Planning and Inference\par
Abstract: This paper applies methodology of Finkelstein and Schoenfeld [Stat. Med. 13 (1994) 1747.] to consider new treatment strategies in a synthetic clinical trial. The methodology is an approach for estimating survival functions as a composite of subdistributions defined by an auxiliary event which is intermediate to the failure...
\par
Keywords: survival; progression; semi-Markov; AIDS\par
Judgment: \textbf{NO} \\
\end{longtable}
\end{scriptsize}

\section{Results}

In this section, we present the results of our analysis. We report the performance of four LLMs, namely ChatGPT 4o mini, DeepSeek R1, DeepSeek V3, and Gemini 2.0 Flash. For each model, we consider three highly cited thresholds (Top 1\%, Top 5\% and Top 10\%) and compute the ACC, TPR and FPR. We also record the number of papers that are predicted as positives as well as their proportions. This helps us to understand the prediction tendency of the LLMs. In addition, we report the total runtime and the associated cost in USD to provide a practical assessment of computational efficiency. Finally, we assess the stability of LLM predictions by comparing repeated outputs under identical prompts.

\subsection{Overall Performance}

We first provide an overview of the computational characteristics of our evaluation. Table~\ref{tab:model_runtime} reports the runtime, the unified cost in USD, and the number of processed samples for each model. For clarity of comparison, we group papers into six five-year periods covering 1991 to 1995, 1996 to 2000, 2001 to 2005, 2006 to 2010, 2011 to 2015, and 2016 to 2020. In addition, we present the sample sizes for each group. These summaries describe the computing cost of large-scale evaluation.

\begin{table}[htbp]
  \centering
  \caption{Model runtime by publication group. Costs are unified to USD. RMB amounts are converted at an assumed rate of 1 USD = 7.20 CNY.}
  \label{tab:model_runtime}
  \resizebox{\textwidth}{!}{
  \begin{tabular}{cccccc}
    \toprule
    \textbf{Publication Group} &
    \textbf{Sample Size} &
    \textbf{DeepSeek V3} &
    \textbf{DeepSeek R1} &
    \textbf{Gemini 2.0 Flash} &
    \textbf{ChatGPT 4o mini} \\
    \midrule
    1991--1995 & 7,221  & 10h38min & 126h40min & 32h5min  & 1h17min \\
    1996--2000 & 9,628  & 15h31min & 111h10min & 45h30min & 1h46min \\
    2001--2005 & 10,145 & 16h13min & 214h50min & 32h50min & 3h22min \\
    2006--2010 & 13,291 & 20h20min & 175h26min & 42h35min & 2h11min \\
    2011--2015 & 17,378 & 32h31min & 239h51min & 91h10min & 2h56min \\
    2016--2020 & 20,530 & 34h26min & 276h40min & 78h25min & 3h49min \\
    \midrule
    \textbf{Total cost (USD)} & 78,193 & 12.27 & 117.50 & 20.30 & 12.73 \\
    \bottomrule
  \end{tabular}}
  \vspace{0.25em}
\end{table}

The sample size column shows a steady increase from 7,221 in 1991 to 1995 to 20,530 in 2016 to 2020, indicating a steady growth in the number of papers over time. Across publication groups, ChatGPT 4o mini is consistently the fastest and often finishes a group in one to three hours, while DeepSeek V3 usually needs three to five times longer, and Gemini 2.0 Flash lies in between with runtimes that increase with group size. DeepSeek R1 is by far the slowest, with runtimes that in some periods reach nearly one hundred times those of DeepSeek V3. In terms of cost, DeepSeek V3 and ChatGPT 4o mini are the least expensive and differ slightly, Gemini 2.0 Flash is higher but still close to them. DeepSeek R1 is much more expensive and far above the other LLMs. Taken together, DeepSeek V3 shows advantages in both time and cost, while DeepSeek R1 lies at the opposite extreme. This contrast may reflect the fact that DeepSeek R1 is designed as a deep reasoning model, which requires substantially more computation \cite{deepseekR1_nature_2025}.

\begin{table}[htbp]
  \centering
  \caption{Predicted positives by publication group and model. 
    The table reports the count and proportion of papers predicted as highly cited across six publication groups for each of the four LLMs.}
    
  \label{tab:positives_by_group}
  \resizebox{\textwidth}{!}{
  \begin{tabular}{
    l
    S[table-format=5.0]
    S[table-format=5.0]  S[table-format=2.1, round-mode=places, round-precision=1]
    S[table-format=5.0]  S[table-format=2.1, round-mode=places, round-precision=1]
    S[table-format=5.0]  S[table-format=2.1, round-mode=places, round-precision=1]
    S[table-format=5.0]  S[table-format=2.1, round-mode=places, round-precision=1]
  }
    \toprule
    \textbf{Publication Group} & \textbf{Sample Size} &
    \multicolumn{2}{c}{\textbf{DeepSeek V3}} &
    \multicolumn{2}{c}{\textbf{DeepSeek R1}} &
    \multicolumn{2}{c}{\textbf{Gemini 2.0 Flash}} &
    \multicolumn{2}{c}{\textbf{ChatGPT 4o mini}} \\
    \cmidrule(lr){3-4}\cmidrule(lr){5-6}\cmidrule(lr){7-8}\cmidrule(lr){9-10}
    & & \textbf{Count} & \textbf{Proportion (\%)} & \textbf{Count} & \textbf{Proportion (\%)} & \textbf{Count} & \textbf{Proportion (\%)} & \textbf{Count} & \textbf{Proportion (\%)} \\
    \midrule
    1991--1995 & 7221  & 1841 & 25.5 & 1409 & 19.5 & 1797 & 24.9 & 2309 & 31.9 \\
    1996--2000 & 9628  & 1946 & 20.2 & 1598 & 16.5 & 2307 & 23.9 & 3365 & 34.9 \\
    2001--2005 & 10145 & 1280 & 12.6 &  787 &  7.7 & 1827 & 18.0 & 2842 & 28.0 \\
    2006--2010 & 13291 & 1907 & 14.3 & 1638 & 12.3 & 2271 & 17.1 & 4942 & 37.2 \\
    2011--2015 & 17378 & 2605 & 15.0 & 2817 & 16.2 & 4400 & 25.3 & 6535 & 37.6 \\
    2016--2020 & 20530 & 5709 & 27.8 & 6395 & 31.1 & 5290 & 25.8 & 10622 & 51.7 \\
    \midrule
    \textbf{Average} & {} & {} & 19.2 & {} & 17.2 & {}& 22.5 & {} & 36.9 \\
    \bottomrule
  \end{tabular}}
\end{table}

\begin{figure}[htbp]
    \centering
    \includegraphics[width=0.9\textwidth]{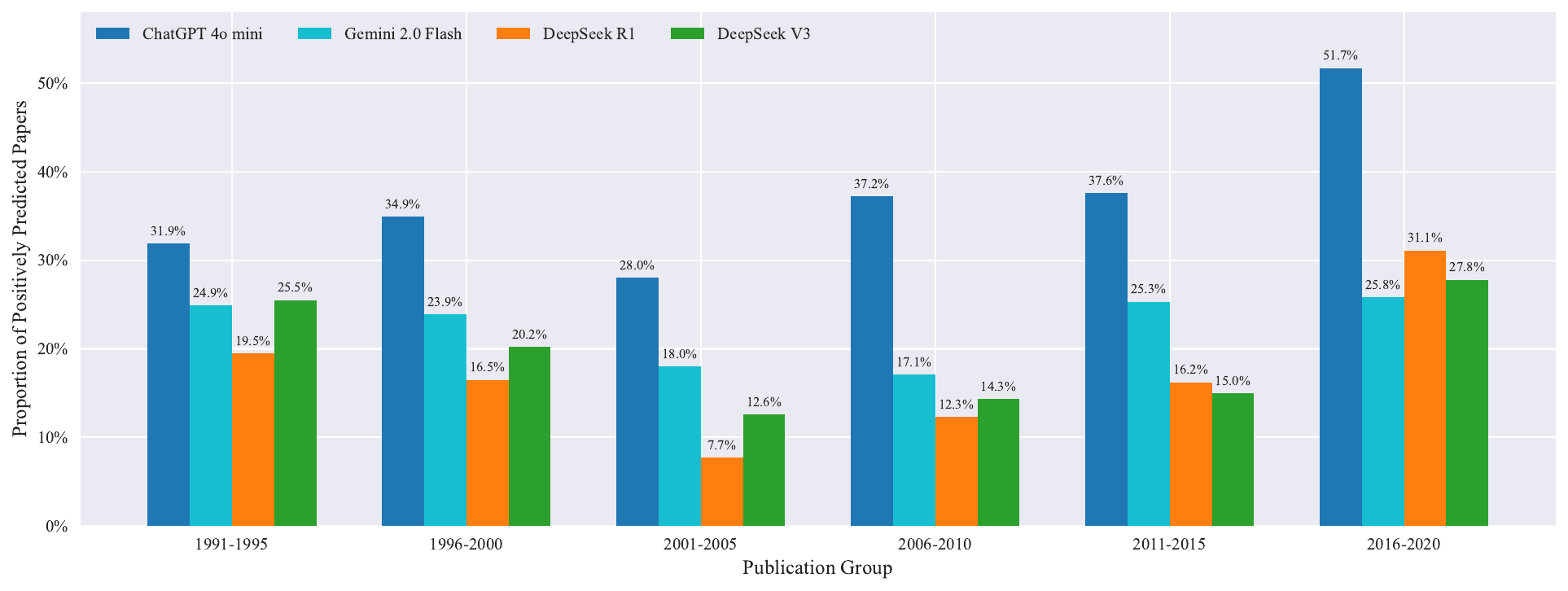}
    \caption{Predicted positive rates across publication groups for ChatGPT 4o mini, Gemini 2.0 Flash, DeepSeek R1 and DeepSeek V3. The figure compares how frequently each model classifies papers as highly cited within each group.}
    \label{fig:share_predicted_positives}
\end{figure}

To understand how strongly each model tends to make positive predictions, we next examine predicted positives across publication groups and LLMs. Table~\ref{tab:positives_by_group} reports the counts and proportions, and Figure~\ref{fig:share_predicted_positives} illustrates how these proportions evolve over time. ChatGPT 4o mini is the model most inclined to make positive predictions, with proportions exceeding 30\% in almost every publication group. In contrast, DeepSeek R1, DeepSeek V3, and Gemini 2.0 Flash are more conservative. Among them, DeepSeek R1 is the most cautious, with an average proportion of only 17.2\%. In addition, all four LLMs assign higher positive proportions in the most recent period. The increase is especially pronounced for ChatGPT 4o mini, which reaches 51.7\% in 2016 to 2020. Even the most conservative DeepSeek V3 increases to around 30\% in this period, indicating a substantially stronger tendency to classify recent papers as highly cited. These patterns suggest that LLMs have an optimistic view of current papers and the research topics they address.

\subsection{Comparison of LLMs}

We next summarize the prediction results under three definitions of highly cited papers, i.e., Top 1\%, Top 5\%, and Top 10\%. Specifically, we report ACC, TPR, and FPR for each LLM and publication group. ACC measures the proportion of correct predictions among all papers, TPR measures the proportion of truly highly cited papers correctly identified as positive, and FPR measures the proportion of non–highly cited papers incorrectly labeled as positive. Detailed results are presented in Table \ref{tab:highly_cited_results}.

\begin{table}[htbp]
  \centering
    \caption{Prediction performance under different definitions of highly cited papers. 
    The table reports ACC, TPR, and FPR for DeepSeek V3, DeepSeek R1, Gemini 2.0 Flash, and ChatGPT 4o mini across six publication groups and three thresholds.}

  \label{tab:highly_cited_results}

 \textbf{(a) Top 1\%} \\[0.3em]
 \resizebox{\textwidth}{!}{
 \begin{tabular}{l *{12}{S[table-format=1.3, round-mode=places, round-precision=3]}}
   \toprule
    \textbf{Publication Group} &
   \multicolumn{3}{c} {\textbf{DeepSeek V3}} &
   \multicolumn{3}{c} {\textbf{DeepSeek R1}} &
   \multicolumn{3}{c} {\textbf{Gemini 2.0 Flash}} &
   \multicolumn{3}{c} {\textbf{ChatGPT 4o mini}} \\
   \cmidrule(lr){2-4}\cmidrule(lr){5-7}\cmidrule(lr){8-10}\cmidrule(lr){11-13}
   &  \textbf{ACC} &  \textbf{TPR} &  \textbf{FPR} &  \textbf{ACC} &  \textbf{TPR} &  \textbf{FPR} &  \textbf{ACC} &  \textbf{TPR} &  \textbf{FPR} &  \textbf{ACC} &  \textbf{TPR} &  \textbf{FPR} \\
   \midrule
   1991--1995 & 0.754 & 0.943 & 0.248 & 0.812 & 0.853 & 0.188 & 0.760 & 0.907 & 0.242 & 0.689 & 0.907 & 0.314 \\
   1996--2000 & 0.805 & 0.883 & 0.195 & 0.839 & 0.737 & 0.160 & 0.768 & 0.883 & 0.233 & 0.658 & 0.869 & 0.344 \\
   2001--2005 & 0.879 & 0.780 & 0.120 & 0.926 & 0.690 & 0.071 & 0.826 & 0.800 & 0.174 & 0.725 & 0.762 & 0.275 \\
   2006--2010 & 0.863 & 0.847 & 0.136 & 0.880 & 0.640 & 0.118 & 0.837 & 0.885 & 0.164 & 0.635 & 0.824 & 0.367 \\
   2011--2015 & 0.856 & 0.778 & 0.144 & 0.841 & 0.676 & 0.157 & 0.755 & 0.901 & 0.247 & 0.631 & 0.852 & 0.371 \\
   2016--2020 & 0.729 & 0.872 & 0.272 & 0.693 & 0.721 & 0.307 & 0.749 & 0.857 & 0.252 & 0.491 & 0.928 & 0.513 \\
   \midrule
   \textbf{Average} & 0.814 & 0.850 & 0.186 & 0.832 & 0.720 & 0.167 & 0.782 & 0.872 & 0.219 & 0.638 & 0.857 & 0.364 \\
   \bottomrule
  \end{tabular}}
  \vspace{1em}

    \textbf{(b) Top 5\%} \\[0.3em]
    \resizebox{\textwidth}{!}{
    \begin{tabular}{l *{12}{S[table-format=1.3, round-mode=places, round-precision=3]}}
      \toprule
    \textbf{Publication Group} &
   \multicolumn{3}{c} {\textbf{DeepSeek V3}} &
   \multicolumn{3}{c} {\textbf{DeepSeek R1}} &
   \multicolumn{3}{c} {\textbf{Gemini 2.0 Flash}} &
   \multicolumn{3}{c} {\textbf{ChatGPT 4o mini}} \\
   \cmidrule(lr){2-4}\cmidrule(lr){5-7}\cmidrule(lr){8-10}\cmidrule(lr){11-13}
   &  \textbf{ACC} &  \textbf{TPR} &  \textbf{FPR} &  \textbf{ACC} &  \textbf{TPR} &  \textbf{FPR} &  \textbf{ACC} &  \textbf{TPR} &  \textbf{FPR} &  \textbf{ACC} &  \textbf{TPR} &  \textbf{FPR} \\
   \midrule
      1991--1995 & 0.783 & 0.879 & 0.222 & 0.833 & 0.780 & 0.164 & 0.783 & 0.816 & 0.219 & 0.713 & 0.824 & 0.293 \\
      1996--2000 & 0.830 & 0.816 & 0.170 & 0.851 & 0.667 & 0.139 & 0.790 & 0.798 & 0.210 & 0.684 & 0.829 & 0.324 \\
      2001--2005 & 0.892 & 0.676 & 0.097 & 0.918 & 0.461 & 0.057 & 0.840 & 0.696 & 0.153 & 0.743 & 0.731 & 0.256 \\
      2006--2010 & 0.879 & 0.721 & 0.113 & 0.884 & 0.572 & 0.099 & 0.853 & 0.739 & 0.141 & 0.658 & 0.796 & 0.349 \\
      2011--2015 & 0.870 & 0.698 & 0.121 & 0.848 & 0.600 & 0.139 & 0.777 & 0.796 & 0.224 & 0.654 & 0.799 & 0.354 \\
      2016--2020 & 0.754 & 0.817 & 0.249 & 0.712 & 0.736 & 0.289 & 0.771 & 0.785 & 0.230 & 0.521 & 0.883 & 0.498 \\
      \midrule
      \textbf{Average} & 0.835 & 0.768 & 0.162 & 0.841 & 0.636 & 0.148 & 0.802 & 0.772 & 0.196 & 0.662 & 0.810 & 0.346 \\
      \bottomrule
    \end{tabular}}    
  \vspace{1em}

    \textbf{(c) Top 10\%} \\[0.3em]
    \resizebox{\textwidth}{!}{
    \begin{tabular}{l *{12}{S[table-format=1.3, round-mode=places, round-precision=3]}}
      \toprule
    \textbf{Publication Group} &
   \multicolumn{3}{c} {\textbf{DeepSeek V3}} &
   \multicolumn{3}{c} {\textbf{DeepSeek R1}} &
   \multicolumn{3}{c} {\textbf{Gemini 2.0 Flash}} &
   \multicolumn{3}{c} {\textbf{ChatGPT 4o mini}} \\
   \cmidrule(lr){2-4}\cmidrule(lr){5-7}\cmidrule(lr){8-10}\cmidrule(lr){11-13}
   &  \textbf{ACC} &  \textbf{TPR} &  \textbf{FPR} &  \textbf{ACC} &  \textbf{TPR} &  \textbf{FPR} &  \textbf{ACC} &  \textbf{TPR} &  \textbf{FPR} &  \textbf{ACC} &  \textbf{TPR} &  \textbf{FPR} \\
   \midrule
      1991--1995 & 0.803 & 0.790 & 0.195 & 0.837 & 0.657 & 0.143 & 0.798 & 0.735 & 0.194 & 0.733 & 0.763 & 0.270 \\
      1996--2000 & 0.844 & 0.730 & 0.143 & 0.851 & 0.583 & 0.119 & 0.807 & 0.731 & 0.184 & 0.704 & 0.765 & 0.303 \\
      2001--2005 & 0.887 & 0.565 & 0.077 & 0.895 & 0.364 & 0.046 & 0.839 & 0.596 & 0.134 & 0.755 & 0.673 & 0.236 \\
      2006--2010 & 0.876 & 0.597 & 0.092 & 0.872 & 0.477 & 0.083 & 0.850 & 0.605 & 0.122 & 0.677 & 0.743 & 0.330 \\
      2011--2015 & 0.871 & 0.601 & 0.099 & 0.844 & 0.529 & 0.121 & 0.791 & 0.717 & 0.201 & 0.673 & 0.743 & 0.335 \\
      2016--2020 & 0.774 & 0.752 & 0.224 & 0.728 & 0.692 & 0.268 & 0.785 & 0.707 & 0.206 & 0.555 & 0.853 & 0.479 \\
      \midrule
      \textbf{Average} & 0.842 & 0.672 & 0.138 & 0.838 & 0.550 & 0.130 & 0.812 & 0.682 & 0.174 & 0.683 & 0.757 & 0.326 \\
      \bottomrule
    \end{tabular}}

\end{table}

We begin by examining performance under the Top 1\% criterion, which represents the most stringent and challenging setting. 
The results, summarized in Table~\ref{tab:highly_cited_results}a, reveal substantial performance heterogeneity across LLMs and publication groups.
Under this criterion, DeepSeek R1 demonstrates the strongest overall performance. 
Its average ACC reaches 0.832, the highest among all models, while also maintaining the lowest FPR at 0.167. 
Notably, the ACC value peaks at 0.926 for papers published between 2001 and 2005, indicating a strong ability to identify extremely highly cited papers while effectively limiting false positives.
DeepSeek V3 follows closely with an average ACC of 0.814. 
Compared with R1, it achieves a higher TPR of 0.850 versus 0.720, at the cost of slightly reduced precision.
Gemini 2.0 Flash occupies an intermediate position, whereas ChatGPT~4o~mini exhibits the weakest performance, with an ACC of 0.638, reflecting a substantially lower capacity to correctly identify highly cited papers under the Top~1\% criterion.
Across all models, performance declines noticeably for papers published between 2016 and 2020.
This decline likely reflects the fact that recent papers have not yet developed the stable citation patterns. Additionally, shifting research trends make it harder for models to use historical data to predict the success of new studies.
We then assess robustness with respect to alternative citation thresholds.
The results for the Top~5\% and Top~10\% criteria, reported in Tables~\ref{tab:highly_cited_results}b and~\ref{tab:highly_cited_results}c, show systematic changes in model behavior as the definition of highly cited papers becomes broader.
As the threshold expands from the Top~1\% to the Top~5\% and Top~10\%, ACC generally improves or remains stable, while TPR declines across all models, reflecting a more conservative identification of positives.
The DeepSeek models exhibit particularly clear adjustments in prediction conservatism, with the FPR of DeepSeek~R1 decreasing from 0.167 to 0.130.
ChatGPT~4o~mini also shows a reduction in false positives but remains the most aggressive model across all settings, suggesting that its broader prediction tendency is persistent rather than threshold-specific.

In summary, the DeepSeek family delivers the strongest overall performance across definitions of highly cited papers and time periods. DeepSeek R1 is the preferred choice with precision as the main priority, especially when the goal is to identify the most exceptional papers while keeping false positives to a minimum. DeepSeek V3, on the other hand, offers the most balanced profile, retaining substantially more true positives than R1 while maintaining lower FPR than ChatGPT 4o mini. Gemini 2.0 Flash provides stable mid-range behavior, whereas ChatGPT 4o mini functions best as a high-TPR tool for broad exploratory screening. In general, the choice of LLMs should depend on the practical needs of the task. In the following prediction exercises, after considering runtime, computational cost, and empirical performance, we ultimately select DeepSeek V3 as the model for forecasting future highly cited papers.

\subsection{The Stability of LLMs}

This section evaluates the stability of LLM predictions. Stability is an important aspect of reliability in LLM-based classification tasks, as models may produce random outputs even when given identical inputs \cite{atil2024llm,gao2025comparisondeepseekllms}.
To quantify stability, we adopt a simple agreement metric. For each paper, the same prompt and metadata are submitted twice. Then, for each publication group, the agreement score is defined as the proportion of cases where the two predictions are identical. Table~\ref{tab:llm_stability} summarizes the agreement scores across different publication groups under the Top~5\% highly cited criterion for the four LLMs considered. It can be seen that the average agreement scores are  high, with all models exceeding 0.92. This indicates that despite potential randomness in generation, the four LLMs behave reliably under repeated runs. Specifically, DeepSeek V3 stands out as the most stable model, achieving an average agreement score of 0.973. Its performance is strong in all publication groups (with agreement scores between 0.962 and 0.980) and remains stable over time. This suggests that DeepSeek V3 maintains stable prediction behaviors regardless of publication year, even though citation patterns differ noticeably between decades.

\begin{table}[htbp]
\centering
\caption{Stability of LLM performance across publication groups under the Top~5\% highly cited criterion. Agreement scores indicate high reproducibility across all models.}
\label{tab:llm_stability}
\resizebox{\textwidth}{!}{
\begin{tabular}{lcccc}
\toprule
\textbf{Publication Group}
& \textbf{DeepSeek V3}
& \textbf{DeepSeek R1}
& \textbf{Gemini 2.0 Flash}
& \textbf{ChatGPT 4o mini} \\
\midrule
1991--1995 & 0.970 & 0.918 & 0.952 & 0.930 \\
1996--2000 & 0.962 & 0.905 & 0.950 & 0.934 \\
2001--2005 & 0.977 & 0.943 & 0.962 & 0.949 \\
2006--2010 & 0.980 & 0.920 & 0.956 & 0.932 \\
2011--2015 & 0.976 & 0.925 & 0.952 & 0.943 \\
2016--2020 & 0.971 & 0.906 & 0.953 & 0.946 \\
\midrule
\textbf{Average} & \textbf{0.973} & \textbf{0.920} & \textbf{0.954} & \textbf{0.939} \\
\bottomrule
\end{tabular}
}
\end{table}

Gemini 2.0 Flash and ChatGPT 4o mini have a second level of stability. Their agreement scores are around 0.95 and neither model shows large variability.
DeepSeek R1 shows the lowest agreement score at 0.92 on average. To be specific, the gap between the highest value, 0.943 in 2001 to 2005, and the lowest value, 0.905 in 2016 to 2020, is nearly four percentage points. This indicates that DeepSeek R1 is more sensitive to input. One possible explanation is that the reasoning-style generation of DeepSeek R1 introduces some unpredictability, making binary classifications slightly less consistent. Together, the results highlight three key points. First, all models demonstrate high reproducibility, indicating that LLM-based classification of highly cited papers is reliable. Second, the differences between models are consistent, with DeepSeek V3 performing best, DeepSeek R1 lagging behind other models in stability, and the other two models falling in between. Third, temporal changes in the dataset have little effect on stability, suggesting that agreement is primarily determined by model design rather than publication year. Overall, the stability analysis indicates that the prediction results presented in the following sections are robust.

\section{Future Research Directions}

This section aims to identify emerging research directions by examining recent publications that are predicted to become highly cited. To this end, we construct a test dataset from \textit{Web of Science}, covering the same 42 journals as in our main analysis and including papers published between 2021 and 2023. At such an early citation horizon, conventional bibliometric indicators provide limited information, and the long-term impact of these papers is not yet observable. Therefore, we rely on our prompted-based prediction to extract early semantic signals associated with future high impact. Specifically, we employ the DeepSeek V3 model to generate impact predictions for each paper based solely on its textual content. In the following subsection, we summarize the prediction results from three perspectives: journal level, topic level, and paper level. We then provide a descriptive analysis of the papers predicted to have high impact to uncover potential research directions.

\subsection{Journal-level Perspective}

We apply our model to 11{,}974 papers published between 2021 and 2023, among which 1{,}682 papers are identified as potentially highly cited candidates, accounting for 14.05\% of the total. The predicted rates are 13.03\% for 2021, 14.28\% for 2022, and 15.12\% for 2023, indicating a relatively stable distribution over time with a upward trend. To examine how these papers are distributed across publishers, we further aggregate the results at the journal level. Table~\ref{tab:top_journals_predicted} reports the journals with the highest frequencies of predicted highly cited papers. This journal-level concentration provides an initial indication of which publishers are more likely to host emerging influential research during the early citation stage. From Table~\ref{tab:top_journals_predicted}, seven journals each publish more than 100 papers predicted to become highly cited over the period 2021–2023. Among them, the {\it Journal of the American Statistical Association} (JASA) ranks first with 426 predicted high-impact papers, followed by the {\it Annals of Statistics} (AoS) with 245 papers. The {\it Journal of Business \& Economic Statistics} (JBES) and the {\it Journal of the Royal Statistical Society: Series B} (JRSS-B) also contribute substantial numbers of papers predicted to have high future impact. Notably, these journals are widely recognized as core publishers in the field of statistics, with JASA, AoS, JRSS-B, and Biometrika commonly regarded as the leading journals in statistics. The year-by-year breakdown further shows that these journals maintain a stable leading position across 2021, 2022, and 2023, indicating a persistent and concentrated distribution of high-potential research at the journal level. This journal-level concentration provides important insights into the identification of future research hdirections. Papers predicted to have high future impact are not evenly distributed across publishers, but instead tend to cluster within a small number of leading journals that define the research frontiers. 
These journals tend to host a disproportionate share of papers predicted to have high future impact, suggesting a strong concentration of early influential research within a small set of leading outlets.

\begin{table}[htbp]
  \centering
  \caption{Top journals by predicted highly cited papers (2021--2023). Results highlight a strong concentration of high-impact research within a few elite statistical journals.}
  \label{tab:top_journals_predicted}
  \footnotesize
  \begin{tabularx}{\textwidth}{@{} >{\raggedright\arraybackslash}X r r r r @{}} 
    \toprule
    \textbf{Publisher (Journal)} & \textbf{Total} & \textbf{2021} & \textbf{2022} & \textbf{2023} \\
    \midrule
    \textit{JOURNAL OF THE AMERICAN STATISTICAL ASSOCIATION} & 426 & 176 & 131 & 119 \\
    \textit{ANNALS OF STATISTICS} & 245 & 92 & 70 & 83 \\
    \textit{JOURNAL OF BUSINESS \& ECONOMIC STATISTICS} & 141 & 60 & 51 & 30 \\
    \textit{JOURNAL OF THE ROYAL STATISTICAL SOCIETY: SERIES B--STATISTICAL METHODOLOGY} & 132 & 39 & 40 & 53 \\
    \textit{ANNALS OF APPLIED STATISTICS} & 111 & 31 & 42 & 38 \\
    \textit{BIOMETRICS} & 111 & 34 & 37 & 40 \\
    \textit{BIOMETRIKA} & 110 & 26 & 50 & 34 \\
    \bottomrule
  \end{tabularx}
\end{table}

To better understand the academic context underlying this concentration, we briefly outline the primary scope of these journals. JASA covers the major areas of statistical science, including statistical theory, methodology, computation, and domain-driven, problem-oriented applications. AoS is widely regarded as the premier outlet for modern statistical theory, emphasizing rigor and foundational methodological contributions. JBES focuses on statistical and econometric methods in economics, finance, and related empirical fields, with a strong applied orientation. JRSS-B emphasizes innovative statistical methodology with broad applicability and substantial potential to influence practice. The {\it Annals of Applied Statistics} (AoAS) primarily publishes methodological advances motivated by substantive problems in applied domains, highlighting the interaction between statistical innovation and real-world scientific challenges. Finally, {\it Biometrics} and {\it Biometrika} are leading journals in biostatistics, focusing on the development of statistical methods for biological, medical, and health sciences, with {\it Biometrika} placing particular emphasis on theoretical depth and general methodological principles. Taken together, these journals constitute a key academic ecosystem in which future high-impact research directions in statistics are likely to emerge and evolve.

\subsection{Topic-level Perspective}

\begin{figure}[!htbp]
  \centering
  \includegraphics[width=0.95\linewidth]{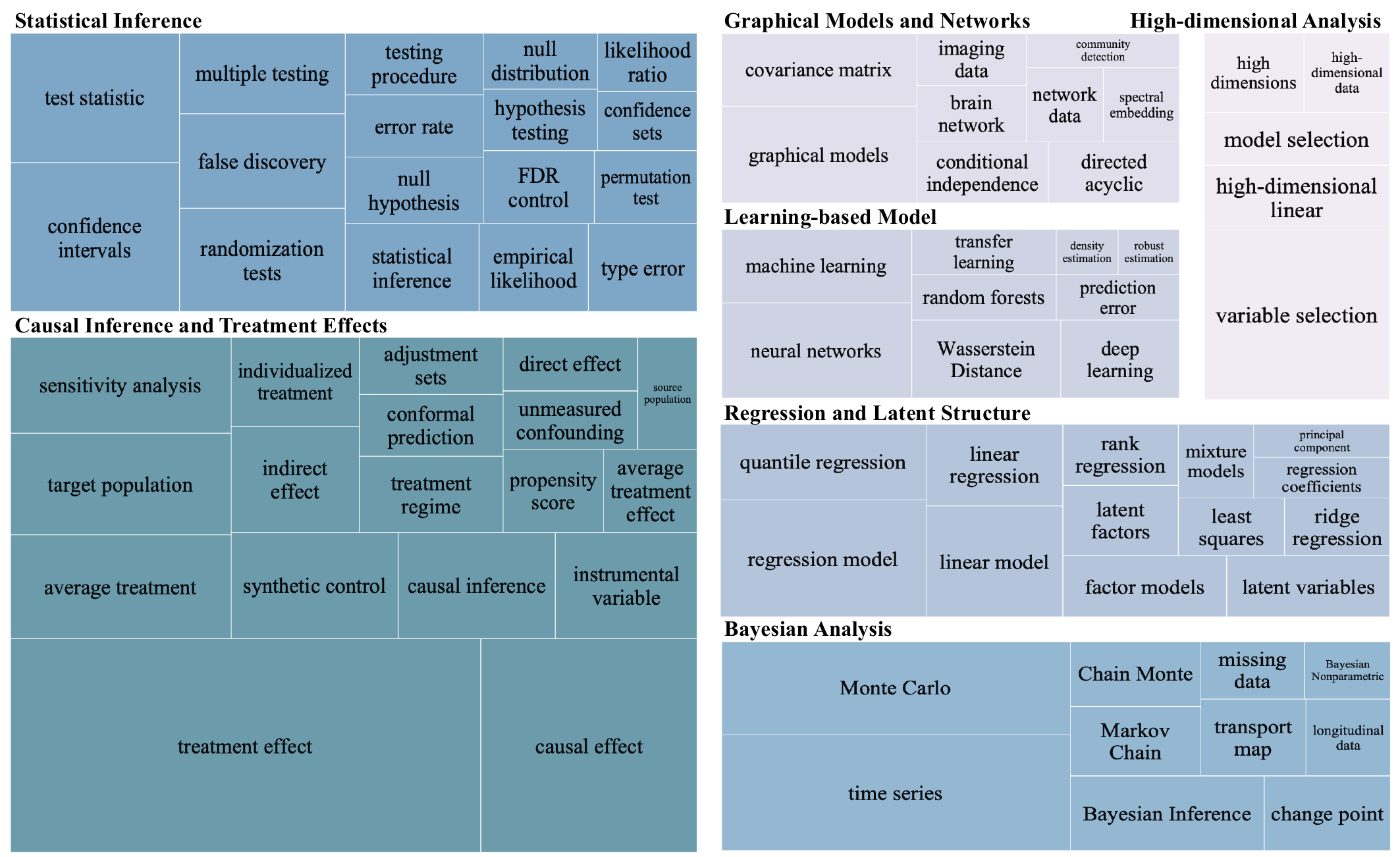}
  \caption{Topic-level distribution of phrases in predicted highly cited papers. The treemap visualizes major research themes and their relative prominence based on phrase frequencies extracted from titles and abstracts.}
  \label{fig:phrase_treemap}
\end{figure}

We examine the distribution of research topics among papers predicted to be highly cited. To characterize topic-level patterns, we apply the TextRank algorithm \cite{mihalcea2004textrank} to extract representative multi-word phrases from titles and abstracts. TextRank is an unsupervised, graph-based method that does not require labeled data, making it well suited for exploratory topic analysis. After extracting phrases, we compute their frequencies based on titles and abstracts, and then manually group phrases that belong to the same methodological domain into broader themes. The resulting topic structure is visualized in Figure~\ref{fig:phrase_treemap} using a treemap representation, where each block corresponds to a phrase and its area reflects the phrase frequency. This visualization provides a compact summary of how predicted highly cited papers are distributed across major methodological domains.

Based on this phrase-level distribution, several dominant topics clearly emerge. The most frequent phrases include \textit{treatment effect} and \textit{causal effect}, followed by general methodological terms such as \textit{time series} and \textit{Monte Carlo}. These phrases are highly prevalent because they correspond to widely applicable problem settings and core analytical tools. Treatment effect analysis is central to empirical research with direct policy relevance and broad cross-disciplinary use. Time series methods form the backbone of empirical analysis in economics, finance, and many applied sciences. Monte Carlo techniques play a fundamental role in simulation-based inference, uncertainty quantification, and modern computational workflows. As a result, research built around these topics tends to attract wide attention and sustain long-term citation impact.

The phrase patterns also form a clear theme structure. We group phrases into seven themes, including Causal Inference and Treatment Effects, Statistical Inference, Bayesian Analysis, Regression and Latent Structure, Learning-based Model, Graphical Models and Networks, and High-dimensional Analysis. To illustrate how these themes correspond to concrete research directions, we highlight representative themes and phrases as follows:

\clearpage

\begin{itemize}
  \item \textit{Causal Inference and Treatment Effects} mainly includes high-frequency phrases such as \textit{treatment effect}, \textit{causal effect}, and \textit{instrumental variable}. This theme is expected to remain highly influential because modern empirical research increasingly requires credible causal claims under complex data collection and imperfect identification conditions. Methods that provide robust identification, transparent assumptions, and policy-relevant estimands tend to diffuse quickly across disciplines.

  \item \textit{Statistical Inference} mainly includes \textit{confidence intervals}, \textit{test statistic}, and \textit{multiple testing}.
  This theme reflects long-run demand for valid uncertainty quantification and error control, which is likely to grow further in large-scale studies and automated pipelines. As research communities emphasize reproducibility and reliability, general inference tools remain foundational and widely citable.

  \item \textit{Bayesian Analysis} includes \textit{bayesian inference}, \textit{Monte Carlo}, and \textit{Markov chain}. The prominence of this theme is forward-looking because Bayesian modeling provides a unified language for uncertainty, hierarchical structures, and decision-making. With increasing computational resources and probabilistic workflows, scalable Bayesian computation and diagnostics are likely to attract sustained attention.

  \item \textit{Regression and Latent Structure} mainly includes \textit{regression model}, \textit{linear model}, and \textit{quantile regression}. This theme remains central because many applied domains still require interpretable models and structured representation learning. Work that connects classical regression principles with modern latent structure modeling is likely to be influential across a wide range of applications.

  \item \textit{Learning-based Model} mainly includes \textit{machine learning}, \textit{neural networks}, and \textit{deep learning}. This theme is likely to generate highly cited contributions when learning-based methods are combined with statistical guarantees, robust evaluation, and uncertainty-aware decision rules. Such integration helps bridge predictive performance with scientific validity, which is increasingly valued in high-stakes domains.

\end{itemize}

Overall, the textual evidence suggests that the predicted highly cited papers mainly focus on several core methodological directions, including causal methodology, principles of statistical inference, Bayesian computation and regression structures, as well as the integration of statistical methodology with machine learning tools. These themes reflect both enduring foundations in statistical theory and emerging methodological directions that are likely to drive future high-impact research.

\subsection{Paper-level Perspective}

To provide a concrete view of the papers flagged as highly cited, we highlight several representative cases as follows:
\begin{itemize}
  \item \textit{What are the Most Important Statistical Ideas of the Past 50 Years?} (Journal of the American Statistical Association) synthesizes foundational developments such as causal inference, simulation-based methodology, and regularization into a comprehensive field-level narrative \citep{gelman2021ideas}.
  \item \textit{Is there a role for statistics in artificial intelligence?} (Advances in Data Analysis and Classification) positions statistics at the core of AI, emphasizing study design, causal reasoning, and methods for assessing uncertainty in data-driven systems \citep{friedrich2022role}.
  \item \textit{To adjust or not to adjust? Estimating the average treatment effect in randomized experiments with missing covariates} (Journal of the American Statistical Association) compares methods for handling missing covariates in randomized experiments under a design-based framework and advocates the missingness-indicator method as a consistent, efficient, and simple approach.  \citep{zhao2024adjust}.
\end{itemize}

All three papers align closely with high-frequency thematic signals surfaced in our text analysis—such as \textit{treatment effects}, \textit{causal inference}, \textit{time series}, and \textit{Monte Carlo}. Rather than focusing solely on domain-specific applications, these works develop broadly applicable methodology, address cross-domain modeling challenges, or provide conceptual synthesis. This pattern suggests that the models tend to associate long-term influence with papers that advance statistical foundations, tackle persistent sources of inferential uncertainty, or extend statistical reasoning into adjacent scientific fields.

\section{Discussion and Conclusions}

\subsection{A Wechat Mini Program }

\begin{figure}[!ht]
    \centering
    \includegraphics[width=1\linewidth]{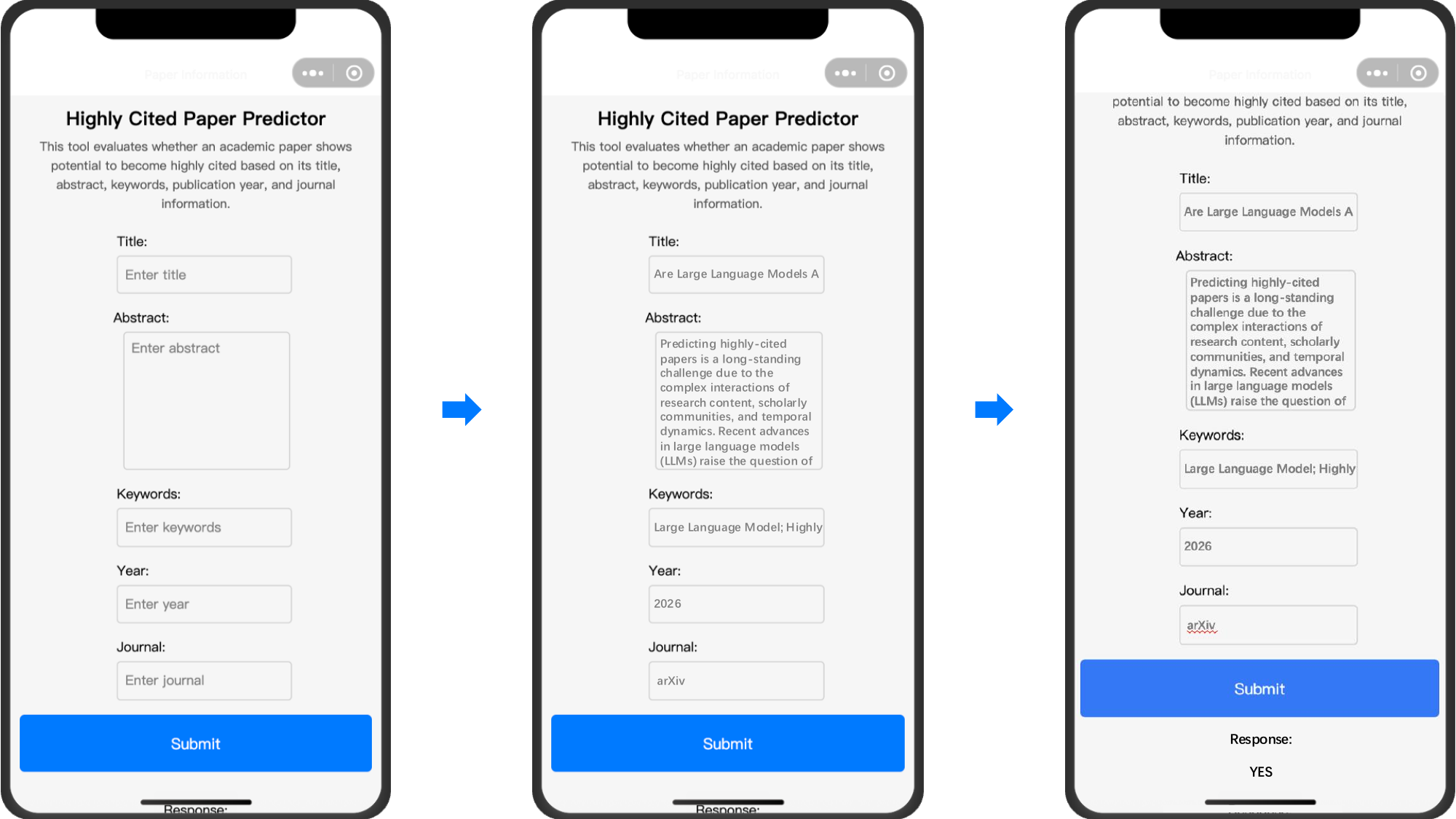}
    \caption{The visual interface of the \textit{STAT Highly-Cited Papers} mini program}
    \label{fig:aicite}
\end{figure}

To facilitate practical use of our proposed methodology, we develope a WeChat mini program named \textit{STAT Highly-Cited Papers}. The tool is designed to provide an accessible interface for researchers to obtain a preliminary assessment of the potential citation impact of their work based on the proposed predictive framework.
The mini program can be accessed by searching for \textit{STAT Highly-Cited Papers} within the WeChat platform. Figure~\ref{fig:aicite} illustrates the visual interface of our \textit{STAT Highly-Cited Papers} mini program.
Users are invited to input basic bibliographic information, including the paper title, abstract, keywords, publication year, and journal. For unpublished manuscripts, such as working papers or preprints, the journal field may be replaced by \textit{arXiv}. Based on the provided information, the system returns a prediction result generated by our model.

We emphasize that the mini program does not retain or store any user-submitted content. All inputs are processed on a per-query basis without memory or data accumulation, and no textual information is used for model updating or retained for future use. As a result, users need not be concerned about data leakage or potential plagiarism risks.
Finally, it is important to note that the output of the mini program should be interpreted as a reference rather than a definitive evaluation. While the proposed model captures early textual and contextual signals associated with future citation outcomes, it cannot replace expert judgment or account for long-term scientific influence shaped by evolving research agendas, communities, and breakthroughs. The tool is intended to support, rather than substitute, human assessment in understanding potential research impact.

\subsection{Limitations}

While the proposed framework demonstrates stable and competitive performance across different publication periods and citation thresholds (as shown in Table~\ref{tab:highly_cited_results}), several limitations should be acknowledged when comparing our results with existing studies summarized in Table~\ref{tab:highly_cited_best_summary}. First, prior work on highly cited paper prediction often relies on task-specific model designs and carefully engineered feature sets. For example, methods based on TabNet, graph neural networks, or ensemble classifiers frequently incorporate rich bibliometric, network, or temporal features, and in some cases achieve very high ACC under narrowly defined prediction targets, such as Top 1\% or fixed citation count thresholds. In contrast, our approach is model-agnostic and text-centered, focusing primarily on semantic information available at or near the time of publication. As a result, while our average ACC levels are competitive, they do not consistently exceed the best reported performance of task-specific models in the literature.

\begin{table}[!ht]
\centering
\scriptsize  
\caption{Summary of Highly-Cited Paper Prediction Studies (Best Model per Paper)}
\label{tab:highly_cited_best_summary}
\begin{tabularx}{\textwidth}{
>{\hsize=0.04\hsize\raggedright\arraybackslash}X
>{\hsize=0.18\hsize\raggedright\arraybackslash}X
>{\hsize=0.20\hsize\raggedright\arraybackslash}X
>{\hsize=0.22\hsize\raggedright\arraybackslash}X
>{\hsize=0.20\hsize\raggedright\arraybackslash}X
>{\hsize=0.16\hsize\raggedright\arraybackslash}X
}
\toprule
\textbf{ID} &
\textbf{Best Model} &
\textbf{Dataset} &
\textbf{Highly-Cited Definition} &
\textbf{Main Features} &
\textbf{Best Performance} \\
\midrule

1 &
TabNet &
Artificial Intelligence &
Top 1\% and Top 5\% \newline
(citation percentile) &
Early citation counts, \newline
bibliometric indicators &
(Top 1\%):\newline
ACC  = 0.98, TPR = 0.59\newline
(Top 5\%):\newline
ACC  = 0.91, TPR = 0.81 \cite{hu2023_ipm} \\

2 & 
LDA + KP-based Supervised Learning & 
Marketing \& MIS & 
Top 25\%, 33\% \& 50\% of highly-cited papers & 
Topic distributions, keyword popularity, bibliometric features & 
6 years after publication(Top 25\%):\newline
Marketing: TPR = 0.65, F1-score = 0.67\newline 
MIS:TPR = 0.73, F1-score = 0.78 \\

3 &
Multinomial Logistic Regression &
Medical systematic reviews &
High-citation trajectory clusters \newline
(10-year citation patterns) &
Early-year citation trajectories &
ACC = 0.76 \newline
Kappa = 0.62 \cite{marques2024ten} \\

4 &
Graph Neural Network &
Machine Learning &
Highly cited papers inferred \newline
from predicted citation trends &
Citation network structure &
F1-score = 0.25--0.60 \newline
(across different datasets) \cite{cummings2020structured} \\

5 &
Bayesian Logistic Regression &
Radiology journals &
Top 10\% by citation counts &
Early citations, \newline
journal-level indicators &
AUC = 0.81 \cite{rosenkrantz2016use} \\

6 &
Rough-set + Ensemble Classifier &
Astronomy and Astrophysics &
Cited more than 275 times &
Content, authorship, \newline
bibliometric features &
ACC = 0.80--0.90 \newline
(under different number of base classifiers) \cite{wang2011mining} \\

\bottomrule
\end{tabularx}
\end{table}

Second, the definition of highly cited varies substantially across studies. As shown in Table~\ref{tab:highly_cited_best_summary}, existing work adopts heterogeneous criteria, ranging from Top 1\% or Top 5\% papers to citation trajectory clusters or absolute citation count cutoffs. Our evaluation explicitly considers multiple percentile-based definitions (Top 1\%, 5\%, and 10\%), which provides robustness but may lead to more conservative performance estimates compared with studies optimized for a single definition. 
Finally, although large language model–based predictors offer strong generalization advantages, they are sensitive to prompt design, model choice, and temporal shifts in scientific writing styles. This dependence, however, also implies a modular and extensible framework, in which advances in large language models can be directly leveraged without fundamental changes to the overall methodology.

Overall, these limitations highlight a fundamental trade-off between generality and specialization. Existing high-performing models often achieve strong results within carefully constrained settings, whereas our framework prioritizes broad applicability, minimal feature engineering, and early-stage assessment. We view these approaches as complementary rather than competing, with our method providing a scalable and transparent baseline for identifying potentially influential research at the time of publication.

\subsection{Conclusions and Future Directions}

This study addresses the question posed in the title: \emph{Are large language models able to predict highly cited papers?} Focusing on statistical publications, our empirical evidence suggests that the answer is affirmative to a meaningful extent. By leveraging large language models with structured prompt design, we show that early-stage textual information—such as titles, abstracts, and keywords—contains informative signals associated with long-term citation impact.
Across multiple publication periods and alternative definitions of highly cited papers, the proposed framework achieves stable and competitive predictive performance. Importantly, these results are obtained without relying on citation networks or post-publication indicators, highlighting the ability of language models to capture substantive and methodological cues at the time of publication. Textual analysis further reveals that papers predicted as highly cited tend to cluster around enduring methodological themes in statistics, including causal inference, uncertainty quantification, classical modeling frameworks, and the integration of statistical and machine learning methods.

Several directions for future research emerge from this work. Incorporating additional structured information, such as collaboration networks or journal characteristics, may further improve predictive accuracy. Modeling temporal dynamics explicitly could enhance robustness to shifts in scientific writing styles and research topics. Finally, as LLMs continue to advance, systematically evaluating how improvements in model capability translate into gains in citation prediction remains an important avenue for future investigation.

\clearpage
\bigskip
\noindent{\bf Funding.} Rui Pan is supported by the National Natural Science Foundation of China (No.~72471254), and by the Program for Innovation Research, the Disciplinary Funds, and the Emerging Interdisciplinary Project of Central University of Finance and Economics.

\clearpage
\bibliographystyle{elsarticle-num}
\bibliography{references}

\clearpage
\appendix
\section{Full Prompts Used in Study}

The structured prompts used in our evaluation consist of five components. Task framing, Evaluation guidelines, and Input constraints and output format are identical across all publication groups and correspond exactly to the specification described in Tables~\ref{tab:prompt-full} and~\ref{tab:prompt-examples}. For these components, we therefore provide only a concise indication of their structure in the tables below, rather than reproducing them in full. By contrast, the Temporal background and Reference examples vary by publication group. These two components are presented in detail for each group. Due to space limitations, all example abstracts shown have been shortened. The full abstracts, together with complete paper metadata, were supplied to the LLMs during all inference runs. The versions in this appendix are solely for presentation clarity and do not reflect the inputs used in the experiments.

\subsection{Prompt of 1991--1995}
\setlength{\LTpre}{8pt}
\setlength{\LTpost}{8pt}

\begin{scriptsize}
\renewcommand{\arraystretch}{1.1}
\begin{longtable}{@{} p{0.18\linewidth} p{0.76\linewidth} @{}} 
\caption{Structured prompt components (1991--1995)}
\label{tab:prompt-1991-1995}\\
\toprule
\textbf{Category} & \textbf{Prompt Content} \\
\midrule
\endfirsthead
\multicolumn{2}{c}{\tablename~\thetable{} -- continued} \\
\toprule
\textbf{Category} & \textbf{Prompt content} \\
\midrule
\endhead
\midrule
\multicolumn{2}{r}{\small Continued on next page} \\
\endfoot
\bottomrule
\endlastfoot

Task framing &
You are an expert in...
\\ \midrule

Evaluation\par guidelines &
\textit{Guidelines for identifying high-citation potential...}\par
\\ \midrule

Temporal\par background &
Papers published in the late 1980s (1986--1990) emerged during a period characterized by the maturation of classical statistical methodology and the early integration of computational tools into empirical analysis. Linear and generalized linear models remained central to statistical practice, while advances in nonlinear regression and time-series modeling expanded the scope of applied inference. Resampling ideas, including early bootstrap methods, began to gain recognition as practical tools for assessing sampling variability in complex settings. In econometrics, substantial attention was devoted to time-series analysis, including autoregressive integrated moving average (ARIMA) models, cointegration concepts, and the treatment of nonstationarity, reflecting growing interest in macroeconomic and financial data. Computational statistics advanced through improved numerical optimization and simulation techniques, although large-scale stochastic simulation methods were still in their infancy. In machine learning and artificial intelligence, research focused primarily on symbolic methods and early connectionist models, with limited diffusion into mainstream statistical practice. Applications in economics, epidemiology, and the social sciences increasingly relied on model-based inference and computational assistance, setting the stage for the more computationally intensive developments that followed in the 1990s. \\ \midrule

Reference\par examples &
Several illustrative examples (three highly cited and three not highly cited) are provided in Table~\ref{tab:prompt-examples-1991-1995}. Abstracts have been shortened for conciseness. \\ \midrule

Input constraints\par and output\par format &
\textit{Operational constraints...}\par
\textit{Paper information for evaluation...}\par
\textit{Required output format...} \\
\end{longtable}
\end{scriptsize}

\begin{scriptsize}
\renewcommand{\arraystretch}{1.1}
\begin{longtable}{@{} p{0.98\linewidth} @{}} 
\caption{Illustrative reference examples (1991--1995)}
\label{tab:prompt-examples-1991-1995}\\
\toprule
\textbf{Examples} \\
\midrule
\endfirsthead
\multicolumn{1}{c}{\tablename~\thetable{} -- continued} \\
\toprule
\textbf{Examples} \\
\midrule
\endhead
\midrule
\multicolumn{1}{r}{\small Continued on next page} \\
\endfoot
\bottomrule
\endlastfoot

Below are several examples illustrating how to distinguish between highly cited and not highly cited papers.\par
\medskip

\textbf{Example 1: Highly Cited}\par
Title: \textit{Controlling the false discovery rate: A practical and powerful approach to multiple testing}\par
Publisher: Journal of the Royal Statistical Society: Series B (Statistical Methodology)\par
Abstract: The common approach to the multiplicity problem calls for controlling the familywise…\par
Keywords: Bonferroni-type procedures; familywise error rate; multiple-comparison procedures; p-values\par
Judgment: \textbf{YES} \\
\addlinespace[1ex]

\textbf{Example 2: Highly Cited}\par
Title: \textit{Operating characteristics of a rank correlation test for publication bias}\par
Publisher: Biometrics\par
Abstract: An adjusted rank correlation test is proposed as a technique for identifying…\par
Keywords: meta-analysis; publication bias; rank correlation\par
Judgment: \textbf{YES} \\
\addlinespace[1ex]

\textbf{Example 3: Highly Cited}\par
Title: \textit{Multivariate adaptive regression splines}\par
Publisher: Annals of Statistics\par
Abstract: A new method is presented for flexible regression modeling of high dimensional data…\par
Keywords: nonparametric multiple regression; multivariable function approximation; statistical learning neural networks; multivariate smoothing; splines; recursive partitioning; AID; CART\par
Judgment: \textbf{YES} \\
\addlinespace[1ex]

\textbf{Example 4: Not Highly Cited}\par
Title: \textit{Assessment of individual and population bioequivalence using the probability that bioavailabilities are similar}\par
Publisher: Biometrics\par
Abstract: In this paper a new method for the assessment of both individual and population…\par
Keywords: bioequivalence; bootstrap; confidence intervals\par
Judgment: \textbf{NO} \\
\addlinespace[1ex]

\textbf{Example 5: Not Highly Cited}\par
Title: \textit{An iterative procedure for the estimation of parameters in a dose-response model}\par
Publisher: Communications in Statistics -- Simulation and Computation\par
Abstract: The least squares estimates of the parameters in the multistage dose-response…\par
Keywords: MSAE regression; multistage dose response model; least squares; nonlinear regression; radiobiology; exponential model\par
Judgment: \textbf{NO} \\
\addlinespace[1ex]

\textbf{Example 6: Not Highly Cited}\par
Title: \textit{A brief review of the role of lattices in rank order statistics}\par
Publisher: Journal of Statistical Planning and Inference\par
Abstract: Partial ordering of the rank order probabilities arising in one-sample, two-sample…\par
Keywords: rank orders; partial ordering; recurrence relations\par
Judgment: \textbf{NO} \\

\end{longtable}
\end{scriptsize}

\subsection{Prompt of 1996--2000}
\setlength{\LTpre}{8pt}
\setlength{\LTpost}{8pt}

\begin{scriptsize}
\renewcommand{\arraystretch}{1.1}
\begin{longtable}{@{} p{0.18\linewidth} p{0.76\linewidth} @{}} 
\caption{Structured prompt components (1996--2000)}
\label{tab:prompt-1996-2000}\\
\toprule
\textbf{Category} & \textbf{Prompt Content} \\
\midrule
\endfirsthead
\multicolumn{2}{c}{\tablename~\thetable{} -- continued} \\
\toprule
\textbf{Category} & \textbf{Prompt content} \\
\midrule
\endhead
\midrule
\multicolumn{2}{r}{\small Continued on next page} \\
\endfoot
\bottomrule
\endlastfoot

Task framing &
You are an expert in...
\\ \midrule

Evaluation\par guidelines &
\textit{Guidelines for identifying high-citation potential...}\par
\\ \midrule

Temporal\par background &
Papers published in the early 1990s (1991--1995) emerged during a period characterized by methodological consolidation and a gradual transition toward computationally intensive approaches. Generalized linear models and nonlinear regression frameworks were widely adopted, while bootstrap and resampling techniques gained prominence for inference under complex data structures. Advances in computational statistics, including the emergence of Markov chain Monte Carlo (MCMC) methods, laid the groundwork for the resurgence of Bayesian inference, enabling applications previously infeasible. Econometrics emphasized cointegration, error-correction models, and unit root testing, reflecting the central role of time-series methods in macroeconomic and financial analysis. Early developments in machine learning, such as neural networks and decision trees, began to attract broader attention, though adoption was limited by computational resources and data availability. Applications in finance, biostatistics, and social sciences increasingly leveraged semi-parametric and simulation-based methods. When assessing contributions from this period, it is critical to evaluate novelty and influence against this backdrop of methodological refinement, expanding computational capacity, and convergence of statistical and algorithmic paradigms.\\ \midrule

Reference\par examples &
Several illustrative examples (three highly cited and three not highly cited) are provided in Table~\ref{tab:prompt-examples-1996-2000}. Abstracts of these examples are shortened for conciseness. \\ \midrule

Input constraints\par and output\par format &
\textit{Operational constraints...}\par
\textit{Paper information for evaluation...}\par
\textit{Required output format...} \\
\end{longtable}
\end{scriptsize}

\begin{scriptsize}
\renewcommand{\arraystretch}{1.1}
\begin{longtable}{@{} p{0.98\linewidth} @{}} 
\caption{Illustrative reference examples (1996--2000)}
\label{tab:prompt-examples-1996-2000}\\
\toprule
\textbf{Examples} \\
\midrule
\endfirsthead
\multicolumn{1}{c}{\tablename~\thetable{} -- continued} \\
\toprule
\textbf{Examples} \\
\midrule
\endhead
\midrule
\multicolumn{1}{r}{\small Continued on next page} \\
\endfoot
\bottomrule
\endlastfoot

Below are several examples illustrating how to distinguish between highly cited and not highly cited papers.\par
\medskip

\textbf{Example 1: Highly Cited}\par
Title: \textit{Regression shrinkage and selection via the Lasso}\par
Publisher: Journal of the Royal Statistical Society: Series B (Statistical Methodology)\par
Abstract: We propose a new method for estimation in linear models…\par
Keywords: quadratic programming; regression; shrinkage; subset selection\par
Judgment: \textbf{YES} \\
\addlinespace[1ex]

\textbf{Example 2: Highly Cited}\par
Title: \textit{A proportional hazards model for the subdistribution of a competing risk}\par
Publisher: Journal of the American Statistical Association\par
Abstract: With explanatory covariates, the standard analysis for competing risks data…\par
Keywords: hazard of subdistribution; martingale; partial likelihood; transformation model\par
Judgment: \textbf{YES} \\
\addlinespace[1ex]

\textbf{Example 3: Highly Cited}\par
Title: \textit{Trim and fill: A simple funnel-plot-based method of testing and adjusting for publication bias in meta-analysis}\par
Publisher: Biometrics\par
Abstract: We study recently developed nonparametric methods for estimating the number of missing studies that might exist in a meta-analysis and the effect that these studies…\par
Keywords: data augmentation; file drawer problem; funnel plots; IQ; malaria; meta-analysis; missing studies; publication bias\par
Judgment: \textbf{YES} \\
\addlinespace[1ex]

\textbf{Example 4: Not Highly Cited}\par
Title: \textit{Limit laws for symmetric k-tensors of regularly varying measures}\par
Publisher: Journal of Multivariate Analysis\par
Abstract: We consider the asymptotics of certain symmetric k-tensors, the vector analogue…\par
Keywords: regularly varying measures; domains of attraction; operator stable laws; symmetric tensors\par
Judgment: \textbf{NO} \\
\addlinespace[1ex]

\textbf{Example 5: Not Highly Cited}\par
Title: \textit{Statistical comparison of axon-scaled neurochemical production}\par
Publisher: Biometrics\par
Abstract: Treatments designed to increase neurochemical levels may also result in increases…\par
Keywords: delta-method approximation; double ratios; independent ratios\par
Judgment: \textbf{NO} \\
\addlinespace[1ex]

\textbf{Example 6: Not Highly Cited}\par
Title: \textit{Nonparametric methods for checking the validity of prior order information}\par
Publisher: Annals of the Institute of Statistical Mathematics\par
Abstract: A large number of statistical procedures have been proposed in the literature to explicitly utilize available information about the ordering of treatment effects at increasing treatment levels…\par
Keywords: distribution-free test; lact-of-fit; ordered null hypothesis; order restricted inferences; partial order\par
Judgment: \textbf{NO} \\

\end{longtable}
\end{scriptsize}

\subsection{Prompt of 2006--2010}
\setlength{\LTpre}{8pt}
\setlength{\LTpost}{8pt}

\begin{scriptsize}
\renewcommand{\arraystretch}{1.1}
\begin{longtable}{@{} p{0.18\linewidth} p{0.76\linewidth} @{}} 
\caption{Structured prompt components (2006--2010)}
\label{tab:prompt-2006-2010}\\
\toprule
\textbf{Category} & \textbf{Prompt Content} \\
\midrule
\endfirsthead
\multicolumn{2}{c}{\tablename~\thetable{} -- continued} \\
\toprule
\textbf{Category} & \textbf{Prompt Content} \\
\midrule
\endhead
\midrule
\multicolumn{2}{r}{\small Continued on next page} \\
\endfoot
\bottomrule
\endlastfoot

Task framing &
You are an expert in...
\\ \midrule

Evaluation\par guidelines &
\textit{Guidelines for identifying high-citation potential...}\par
\\ \midrule

Temporal\par background &
These papers were published around the early 2000s (2001--2005), a period shaped by methodological innovation and expanding computational resources. In Statistics, advances in high-dimensional data analysis, penalized regression (e.g., LASSO), and Bayesian hierarchical modeling enabled models for increasingly complex data. Econometrics progressed in dynamic panel estimation, treatment-effect analysis, and structural modeling, reflecting demand for robust causal inference. In Machine Learning, SVMs, kernel methods, and early ensembles (boosting, bagging) gained prominence. Computational statistics benefited from improvements in MCMC and EM extensions, expanding feasible Bayesian and latent-variable models. Applications accelerated in bioinformatics (post--Human Genome Project) and in NLP/IR with web-scale data.  \\ \midrule

Reference\par examples &
Several illustrative examples (three highly cited and three not highly cited) are provided in Table~\ref{tab:prompt-examples-2006-2010}. Abstracts have been shortened for conciseness. \\ \midrule

Input constraints\par and output\par format &
\textit{Operational constraints...}\par
\textit{Paper information for evaluation...}\par
\textit{Required output format...} \\
\end{longtable}
\end{scriptsize}

\begin{scriptsize}
\renewcommand{\arraystretch}{1.1}
\begin{longtable}{@{} p{0.98\linewidth} @{}} 
\caption{Illustrative reference examples (2006--2010)}
\label{tab:prompt-examples-2006-2010}\\
\toprule
\textbf{Examples} \\
\midrule
\endfirsthead
\multicolumn{1}{c}{\tablename~\thetable{} -- continued} \\
\toprule
\textbf{Examples} \\
\midrule
\endhead
\midrule
\multicolumn{1}{r}{\small Continued on next page} \\
\endfoot
\bottomrule
\endlastfoot

Below are several examples illustrating how to distinguish between highly cited and not highly cited papers.\par
\medskip

\textbf{Example 1: Highly Cited}\par
Title: \textit{Regularization Paths for Generalized Linear Models via Coordinate Descent}\par
Publisher: Journal of Statistical Software\par
Abstract: We develop fast algorithms for estimation of generalized linear models…\par
Keywords: lasso; elastic net; logistic regression; l(1) penalty; regularization path; coordinate-descent\par
Judgment: \textbf{YES} \\
\addlinespace[1ex]

\textbf{Example 2: Highly Cited}\par
Title: \textit{A tutorial on spectral clustering}\par
Publisher: Statistics and Computing\par
Abstract: In recent years, spectral clustering has become one of the most popular modern…\par
Keywords: spectral clustering; graph Laplacian\par
Judgment: \textbf{YES} \\
\addlinespace[1ex]

\textbf{Example 3: Highly Cited}\par
Title: \textit{The adaptive lasso and its oracle properties}\par
Publisher: Journal of the American Statistical Association\par
Abstract: The lasso is a popular technique for simultaneous estimation and variable selection…\par
Keywords: asymptotic normality; lasso; minimax; oracle inequality; oracle procedure; variable selection\par
Judgment: \textbf{YES} \\
\addlinespace[1ex]

\textbf{Example 4: Not Highly Cited}\par
Title: \textit{Comparison of Nonparametric Components in Two Partially Linear Models}\par
Publisher: Communications in Statistics-Theory and Methods\par
Abstract: In this article, we are concerned with whether the nonparametric functions are parallel from two partial linear models, and propose a test statistic to check the difference of the two functions…\par
Keywords: Generalized likelihood ratio test; Local linear estimation; Moment method; Partially linear models; Two stage method\par
Judgment: \textbf{NO} \\
\addlinespace[1ex]

\textbf{Example 5: Not Highly Cited}\par
Title: \textit{Testing the Homogeneity of Two Survival Functions Against a Mixture Alternative Based on Censored Data}\par
Publisher: Communications in Statistics-Simulation and Computation\par
Abstract: When the survival distribution in a treatment group is a mixture of two distributions of the same family, traditional parametric methods that ignore the existence of mixture components or the nonparametric methods may not be very powerful…\par
Keywords: Censored data; Hypothesis testing; Mixture model; Survival functions\par
Judgment: \textbf{NO} \\
\addlinespace[1ex]

\textbf{Example 6: Not Highly Cited}\par
Title: \textit{Sequential Analysis of Longitudinal Data in a Prospective Nested Case-Control Study}\par
Publisher: Biometrics\par
Abstract: The nested case-control design is a relatively new type of observational study whereby a case-control approach is employed within an established cohort…\par
Keywords: Group sequential test; Logistic regression; Longitudinal data; Nested case-control design; Sequential sampling; Stopping time\par
Judgment: \textbf{NO} \\

\end{longtable}
\end{scriptsize}

\subsection{Prompt of 2011--2015}
\setlength{\LTpre}{8pt}
\setlength{\LTpost}{8pt}

\begin{scriptsize}
\renewcommand{\arraystretch}{1.1}
\begin{longtable}{@{} p{0.18\linewidth} p{0.76\linewidth} @{}} 
\caption{Structured prompt components (2011--2015)}
\label{tab:prompt-2011-2015}\\
\toprule
\textbf{Category} & \textbf{Prompt Content} \\
\midrule
\endfirsthead
\multicolumn{2}{c}{\tablename~\thetable{} -- continued} \\
\toprule
\textbf{Category} & \textbf{Prompt Content} \\
\midrule
\endhead
\midrule
\multicolumn{2}{r}{\small Continued on next page} \\
\endfoot
\bottomrule
\endlastfoot

Task framing &
You are an expert in...
\\ \midrule

Evaluation\par guidelines &
\textit{Guidelines for identifying high-citation potential...}\par
\\ \midrule

Temporal\par background &
Papers published in the mid-to-late 2000s (2006--2010) emerged during a period marked by rapid advances in high-dimensional modeling and computational scalability. Sparse modeling and variable selection became central in Statistics, with penalization approaches such as LASSO, SCAD, and elastic net widely applied to complex datasets. Econometrics emphasized causal inference, with difference-in-differences, propensity score matching, and instrumental-variable methods increasingly used in policy and labor applications, while dynamic panel models and nonparametric tools expanded empirical capabilities. Machine Learning entered a transformative stage, with ensemble methods such as random forests maturing, kernel methods becoming more refined, and early neural network research laying groundwork for later breakthroughs. Computational statistics was shaped by scalable MCMC, variational inference, and improved optimization techniques, enabling hierarchical and latent-variable models. Bioinformatics and genomics drove methodological innovation, while finance, marketing, and social sciences adopted data-intensive approaches. Evaluating contributions from this period requires considering methodological refinement, computational scalability, and the early convergence of statistical and algorithmic traditions. \\ \midrule

Reference\par examples &
Several illustrative examples (three highly cited and three not highly cited) are provided in Table~\ref{tab:prompt-examples-2011-2015}. Abstracts have been shortened for conciseness. \\ \midrule

Input constraints\par and output\par format &
\textit{Operational constraints...}\par
\textit{Paper information for evaluation...}\par
\textit{Required output format...} \\
\end{longtable}
\end{scriptsize}

\begin{scriptsize}
\renewcommand{\arraystretch}{1.1}
\begin{longtable}{@{} p{0.98\linewidth} @{}} 
\caption{Illustrative reference examples (2011--2015)}
\label{tab:prompt-examples-2011-2015}\\
\toprule
\textbf{Examples} \\
\midrule
\endfirsthead
\multicolumn{1}{c}{\tablename~\thetable{} -- continued}\\
\toprule
\textbf{Examples}\\
\midrule
\endhead
\midrule
\multicolumn{1}{r}{\small Continued on next page}\\
\endfoot
\bottomrule
\endlastfoot

Below are several examples illustrating how to distinguish between highly cited and not highly cited papers.\par
\medskip

\textbf{Example 1: Highly Cited}\par
Title: \textit{Fitting Linear Mixed-Effects Models Using lme4}\par
Publisher: Journal of Statistical Software\par
Abstract: Maximum likelihood or restricted maximum likelihood (REML) estimates of the parameters in linear mixed-effects models can be determined using the lmer function in the lme4 package for R…\par
Keywords: sparse matrix methods; linear mixed models; penalized least squares; Cholesky decomposition\par
Judgment: \textbf{YES} \\
\addlinespace[1ex]

\textbf{Example 2: Highly Cited}\par
Title: \textit{Fast stable restricted maximum likelihood and marginal likelihood estimation of semiparametric generalized linear models}\par
Publisher: Journal of the Royal Statistical Society Series B-Statistical Methodology\par
Abstract: Recent work by Reiss and Ogden provides a theoretical basis for sometimes preferring restricted maximum likelihood (REML) to generalized cross-validation (GCV) for smoothing parameter selection in semiparametric regression…\par
Keywords: Adaptive smoothing; Generalized additive mixed model; Generalized additive model; Generalized cross-validation; Marginal likelihood; Model selection; Penalized generalized linear model; Penalized regression splines; Restricted maximum likelihood; Scalar on function regression; Stable computation\par
Judgment: \textbf{YES} \\
\addlinespace[1ex]

\textbf{Example 3: Highly Cited}\par
Title: \textit{Robust Inference With Multiway Clustering}\par
Publisher: Journal of Business \& Economic Statistics\par
Abstract: In this article we propose a variance estimator for the OLS estimator as well as for nonlinear estimators such as logit, probit, and GMM…\par
Keywords: Cluster-robust standard errors; Two-way clustering\par
Judgment: \textbf{YES} \\
\addlinespace[1ex]

\textbf{Example 4: Not Highly Cited}\par
Title: \textit{Spurious Regressions in Time Series with Long Memory}\par
Publisher: Communications in Statistics-Theory and Methods\par
Abstract: This article studies the asymptotic properties of least squares estimators and related test statistics in some spurious regression models that are generated by stationary or nonstationary fractionally integrated processes…\par
Keywords: Linear process; Long memory; Spurious regression; Self-normalized sums\par
Judgment: \textbf{NO} \\
\addlinespace[1ex]

\textbf{Example 5: Not Highly Cited}\par
Title: \textit{Simultaneous estimation of the locations and effects of multiple disease loci in case-control studies}\par
Publisher: Biostatistics\par
Abstract: The genetic basis of complex diseases often involves multiple causative loci…\par
Keywords: Association analysis; Case-control design; Generalized estimating equations; Linkage disequilibrium; Multi-locus\par
Judgment: \textbf{NO} \\
\addlinespace[1ex]

\textbf{Example 6: Not Highly Cited}\par
Title: \textit{Simultaneous large deviations for the shape of Young diagrams associated with random words}\par
Publisher: Bernoulli\par
Abstract: We investigate the large deviations of the shape of the random RSK Young diagrams associated with a random word of size n whose letters are independently drawn from an alphabet of size m = m(n)…\par
Keywords: large deviations; longest increasing subsequence; random matrices; random words; strong approximation; Young diagrams\par
Judgment: \textbf{NO} \\

\end{longtable}
\end{scriptsize}

\subsection{Prompt of 2016--2020}
\setlength{\LTpre}{8pt}
\setlength{\LTpost}{8pt}

\begin{scriptsize}
\renewcommand{\arraystretch}{1.1}
\begin{longtable}{@{} p{0.18\linewidth} p{0.76\linewidth} @{}} 
\caption{Structured prompt components (2016--2020)}
\label{tab:prompt-2016-2020}\\
\toprule
\textbf{Category} & \textbf{Prompt Content} \\
\midrule
\endfirsthead
\multicolumn{2}{c}{\tablename~\thetable{} -- continued} \\
\toprule
\textbf{Category} & \textbf{Prompt Content} \\
\midrule
\endhead
\midrule
\multicolumn{2}{r}{\small Continued on next page} \\
\endfoot
\bottomrule
\endlastfoot

Task framing &
You are an expert in...
\\ \midrule

Evaluation\par guidelines &
\textit{Guidelines for identifying high-citation potential...}\par
\\ \midrule

Temporal\par background &
Papers published in the first half of the 2010s (2011--2015) emerged during a period marked by the accelerated adoption of data-driven approaches across Statistics, Econometrics, and Machine Learning. In Statistics, scalable inference for high-dimensional data became mainstream, with penalized likelihood, sparse graphical models, and Bayesian shrinkage priors widely adopted. Econometrics consolidated causal inference frameworks, including synthetic control, regression discontinuity, and advanced instrumental-variable techniques, reflecting increasing focus on treatment effects and policy evaluation with observational data. Machine Learning underwent a transformative phase with deep learning, driven by breakthroughs in convolutional and recurrent architectures, GPU computation, and large labeled datasets, enabling rapid progress in image recognition, speech, and NLP. Computational statistics advanced in parallel through stochastic variational inference, scalable MCMC, and distributed optimization, making hierarchical and Bayesian models feasible at new scales. Applications expanded across genomics, personalized medicine, finance, marketing, and social networks, demanding methods capable of integrating heterogeneous, high-dimensional, and unstructured data. Evaluating contributions from this period requires considering methodological scalability, causal rigor, and the deep learning revolution that redefined research trajectories across disciplines. \\ \midrule

Reference\par examples &
Several illustrative examples (three highly cited and three not highly cited) are provided in Table~\ref{tab:prompt-examples-2016-2020}. Abstracts have been shortened for conciseness. \\ \midrule

Input constraints\par and output\par format &
\textit{Operational constraints...}\par
\textit{Paper information for evaluation...}\par
\textit{Required output format...} \\
\end{longtable}
\end{scriptsize}

\begin{scriptsize}
\renewcommand{\arraystretch}{1.1}
\begin{longtable}{@{} p{0.98\linewidth} @{}} 
\caption{Illustrative reference examples (2016--2020)}
\label{tab:prompt-examples-2016-2020}\\
\toprule
\textbf{Examples} \\
\midrule
\endfirsthead
\multicolumn{1}{c}{\tablename~\thetable{} -- continued} \\
\toprule
\textbf{Examples} \\
\midrule
\endhead
\midrule
\multicolumn{1}{r}{\small Continued on next page} \\
\endfoot
\bottomrule
\endlastfoot

Below are several examples illustrating how to distinguish between highly cited and not highly cited papers.\par
\medskip

\textbf{Example 1: Highly Cited}\par
Title: \textit{lmerTest Package: Tests in Linear Mixed Effects Models}\par
Publisher: Journal of Statistical Software\par
Abstract: One of the frequent questions by users of the mixed model function lmer of the lme4 package has been: How can I get p values for the F and t tests for objects returned by lmer?…\par
Keywords: denominator degree of freedom; Satterthwaite's approximation; ANOVA; R; linear mixed effects models; lme4\par
Judgment: \textbf{YES} \\
\addlinespace[1ex]

\textbf{Example 2: Highly Cited}\par
Title: \textit{Unobservable Selection and Coefficient Stability: Theory and Evidence}\par
Publisher: Journal of Business \& Economic Statistics\par
Abstract: A common approach to evaluating robustness to omitted variable bias is to observe coefficient movements after inclusion of controls…\par
Keywords: Coefficient stability; Selection; Omitted variable bias\par
Judgment: \textbf{YES} \\
\addlinespace[1ex]

\textbf{Example 3: Highly Cited}\par
Title: \textit{Practical Bayesian model evaluation using leave-one-out cross-validation and WAIC}\par
Publisher: Statistics and Computing\par
Abstract: Leave-one-out cross-validation (LOO) and the widely applicable information criterion (WAIC) are methods for estimating pointwise out-of-sample prediction accuracy from a fitted Bayesian model using the log-likelihood evaluated at the posterior simulations of the parameter values…\par
Keywords: Bayesian computation; Leave-one-out cross-validation (LOO); K-fold cross-validation; Widely applicable information criterion (WAIC); Stan; Pareto smoothed importance sampling (PSIS)\par
Judgment: \textbf{YES} \\
\addlinespace[1ex]

\textbf{Example 4: Not Highly Cited}\par
Title: \textit{The determination of biosimilarity margin and the assessment of biosimilarity for an -arm parallel design}\par
Publisher: Communications in Statistics-Theory and Methods\par
Abstract: One of the key issues in biosimilar phase III clinical trials is the determination of biosimilarity margins…\par
Keywords: Equivalence trial; random effect model; multiple arm parallel design; biological product; biosimilar product\par
Judgment: \textbf{NO} \\
\addlinespace[1ex]

\textbf{Example 5: Not Highly Cited}\par
Title: \textit{The particle filter based on random number searching algorithm for parameter estimation}\par
Publisher: Communications in Statistics-Simulation and Computation\par
Abstract: This article addresses the issue of parameter estimation in linear system in the presence of Gaussian noises…\par
Keywords: Linear Gaussian system; Parameter estimation; Particle filter algorithm; Random number searching algorithm\par
Judgment: \textbf{NO} \\
\addlinespace[1ex]

\textbf{Example 6: Not Highly Cited}\par
Title: \textit{Automatic tagging with existing and novel tags}\par
Publisher: Biometrika\par
Abstract: Automatic tagging by key words and phrases is important in multi-label classification of a document…\par
Keywords: Alternating direction method of multipliers; Large margin; Multi-label classification; Scalability; Social bookmarking system; Text mining\par
Judgment: \textbf{NO} \\

\end{longtable}
\end{scriptsize}

\end{document}